\documentclass[numberedappendix,twocolumn]{emulateapj}
\newcommand{\ngal}{N$_{gal}$}

\newcommand{\ngrtwo}{N$_{gal}^{200}$}
\newcommand{\rtwo}{R$_{200}$}

\newcommand{\cgcf}{BGVCF}

\newcommand{\msun}{M$_\sun$}

\newcommand{\kms}{km s$^{-1}$}
\newcommand{\up}{$u$}
\newcommand{\gp}{$g$}
\newcommand{\rp}{$r$}
\newcommand{\ip}{$i$}
\newcommand{\zp}{$z$}

\newcommand{\sigv}{$\sigma$}

\newcommand{\lo}{$L_{opt}$}
\newcommand{\lortwo}{$L_{opt}^{200}$}

\newcommand{\lstar}{L$_{*}$}

\shorttitle{THE CLUSTER VELOCITY-DISPERSION--RICHNESS RELATION}
\shortauthors{BECKER ET AL.}
\begin{document}
\title{The Mean and Scatter of the velocity-dispersion--optical
  richness relation for maxBCG galaxy clusters}
\author{M. R. Becker\altaffilmark{1}, T.A. McKay\altaffilmark{1,2},
  B. Koester\altaffilmark{3}, R. H. Wechsler\altaffilmark{4}, 
  E. Rozo\altaffilmark{5}, A. Evrard\altaffilmark{1,2,6}, 
  D. Johnston\altaffilmark{7}, E. Sheldon\altaffilmark{8},
  J. Annis\altaffilmark{9}, E. Lau\altaffilmark{3},
  R. Nichol\altaffilmark{10}, and C. Miller\altaffilmark{2}}
\altaffiltext{1}{Department of Physics, University of Michigan,
    Ann Arbor, MI 48109}
\altaffiltext{2}{Department of Astronomy, University of Michigan, Ann
  Arbor, MI, 48109}
\altaffiltext{3}{Department of Astronomy and Astrophysics, University
  of Chicago, Chicago, IL, 60637}
\altaffiltext{4}{Kavli Institute for Particle Astrophysics and
  Cosmology, Physics Department, and Stanford Linear Accelerator
  Center, Stanford University, Stanford, CA 94305}
\altaffiltext{5}{CCAPP, The Ohio State University, Columbus, OH 43210}
\altaffiltext{6}{Michigan Center for Theoretical Physics, University
  of Michigan, Ann Arbor, MI 48109}
\altaffiltext{7}{Jet Propulsion Laboratory, Caltech, Pasadena, CA, 91109}
\altaffiltext{8}{Department of Physics, New York University, 4
  Washington Place, New York, NY 10003}
\altaffiltext{9}{Fermi National Accelerator Laboratory, Batavia, IL, 60510}
\altaffiltext{10}{Portsmouth University, Portsmouth England}

\begin{abstract}
  The distribution of galaxies in position and velocity around the
  centers of galaxy clusters encodes important information about
  cluster mass and structure. Using the maxBCG galaxy cluster catalog
  identified from imaging data obtained in the Sloan Digital Sky
  Survey, we study the BCG--galaxy velocity correlation function.  By
  modeling its non-Gaussianity, we measure the mean and scatter in velocity
  dispersion at fixed richness. The mean velocity dispersion increases
  from $202\pm10$ \kms\ for small groups to more than $854\pm102$ \kms\ for large
  clusters. We show the scatter to be at most
  $40.5\pm3.5$\%, declining to $14.9\pm9.4$\% in the richest bins.  We
  test our methods in the C4 cluster catalog, a
  spectroscopic cluster catalog produced from the Sloan Digital Sky
  Survey DR2 spectroscopic sample, and in mock galaxy catalogs
  constructed from N-body simulations.  Our methods are robust,
  measuring the scatter to well within one-sigma of the true value,
  and the mean to within 10\%, in
  the mock catalogs.  By convolving the scatter in velocity dispersion
  at fixed richness with
  the observed richness space density function, we measure the
  velocity dispersion function of the maxBCG galaxy clusters.
  Although velocity dispersion
  and richness do not form a true mass--observable relation, the
  relationship between velocity dispersion and mass is theoretically well
  characterized and has low scatter.  Thus our results provide a key
  link between theory and observations up to the velocity bias between
  dark matter and galaxies.
\end{abstract}

\section{Introduction}
Galaxy clusters play an important role in observations of the
large-scale structure of the Universe. As dramatically non-linear
features in the matter distribution, they stand out as individually
identifiable objects, whose abundant galaxies and hot X-ray emitting
gas provide a rich variety of observable properties. Clusters can be
identified by their galaxy content
\citep{2003ApJS..148..243B,2005ApJS..157....1G,2005AJ....130..968M,2005ApJ...625....6G,
  2006ApJS..167....1B,2007astro.ph..1265K,2007astro.ph..1268K},
their thermal X-ray emission
\citep{1998ApJ...492L..21R,2000ApJS..129..435B,2004astro.ph..3357P},
the Sunyaev-Zeldovich  decrement they
produce in the microwave background signal
\citep{2000ApJ...539...39G,2005MNRAS.359...16L}, or the weak lensing
signature they produce in the shapes of
distant background galaxies \citep{2006ApJ...643..128W}.  Each of these
identification methods also produces proxies for mass: e.g, the number
of galaxies, total stellar luminosity, galaxy velocity dispersion,
X-ray luminosity and temperature, or SZ and weak lensing profiles.

Simulations of the formation and evolution of large-scale structure
through gravitational collapse provide us with rich predictions for
the expected matter distribution within a given cosmology
\citep{2002ApJ...573....7E,2005Natur.435..629S}. These predictions include not only
first-order features, like the halo mass function n(M,z), but
higher-order correlations as well, like the precise way in which
galaxy clusters are themselves clustered as a function of mass.
Comparisons of these theoretical predictions to the observed Universe
provide an excellent opportunity to test our understanding of
cosmology and the formation of large-scale structure.  The weak point
in this chain is that simulations most reliably predict the dark
matter distribution, while observations are most directly sensitive to
luminous galaxies and gas.  Connections between observable properties
and theoretical predictions for dark matter have often been made
through simplifying assumptions that are hard to justify \textit{a
  priori}.

Progress toward solving this problem has been made by the
construction of various mass--observable scaling relations, which are based on
combinations of theoretical predictions and observational
measurements
\citep{2002ApJ...577..569L,2006ApJ...653..954D,2006ApJ...648..956S}.
However, knowledge of the
mean mass at a fixed value of the observable is not sufficient to extract precise
cosmological constraints given the exponential shape of the halo mass
function \citep{2004PhRvD..70d3504L,2005PhRvD..72d3006L}.  To perform
precision cosmology, we must
understand the scatter in the mass--observable relations as well.

Recently, \citet{2006ApJ...648..956S} measured the scatter in the
temperature-luminosity relation for X-ray selected galaxy clusters and
used it to infer the scatter in the mass--luminosity relationship.
Unfortunately, there are relatively few measurements of the scatter in
any mass--observable relationship in the optical. (An exception is an
early observation of scatter in the velocity-dispersion--richness
relationship for a small sample of massive clusters by \citealt{1996A&A...310...31M}.)
For optically selected clusters, the scatter is usually
included as a parameter in the analysis
\citep[e.g.][]{2007ApJ...655..128G,2007astro.ph..3574R,2007astro.ph..3571R}.

The primary goal of this work is to develop a method to estimate both
the mean and scatter in the cluster velocity-dispersion--richness
relation. This comparison between two observable quantities can be made
without reference to structure formation theory. The method developed is
applied to the SDSS maxBCG cluster catalog- a photometrically selected
catalog with extensive spectroscopic follow-up. These methods are tested
extensively with both the C4 catalog \citep{2005AJ....130..968M}, a
smaller spectroscopically-selected sample of
clusters, and new mock catalogs generated by combining N-body simulations
with a prescription for galaxy population (Wechsler et al. 2007, in preparation).

Ultimately, we aim to connect richness to mass through
measurements of velocity dispersion. While the link between dark
matter velocity dispersion and mass is known from N-body simulations to
have very small scatter \citep{2007astro.ph..2241E}, the relationship
between galaxy
and dark matter velocity dispersion (the velocity bias) remains
uncertain and will require additional study. Constraints on the
normalization and scatter of the total mass-richness relation obtained
by this method are thus limited by uncertainty in the velocity bias. 

An outline of the SDSS data and simulations used in this work is
presented in \S \ref{sec:data_and_sims}. In \S \ref{sec:maxbcg}, we
provide a brief overview of the maxBCG cluster finding algorithm and
the properties of the detected cluster sample.  We will focus in
this paper on measurements of the BCG--galaxy velocity correlation
function (\cgcf), which is introduced along with the various fitting
methods we employ in \S
\ref{sec:observables}.  Section \ref{sec:nongauss} presents a new
method for understanding the scatter in the optical richness-velocity dispersion
relation and the computation of the velocity dispersion function for
the maxBCG clusters.  Section \ref{sec:tests} presents measurements of
the \cgcf\ as a function of various cluster
properties.  We connect our velocity dispersion measurements to mass
in \S \ref{sec:massscale}.  Finally, we conclude and discuss future
directions in \S \ref{sec:conclusions}.

\section{SDSS Data and Mock Catalogs}\label{sec:data_and_sims}
\subsection{SDSS Data} 
Data for this study are drawn from the SDSS
\citep{2000AJ....120.1579Y,2004AJ....128..502A,2005AJ....129.1755A,2006ApJS..162...38A},
a combined imaging and spectroscopic survey of $\sim 10^4$ deg$^2$ in
the North Galactic Cap, and a smaller, deeper region in the South.
The imaging survey is carried out in drift-scan mode in the five SDSS
filters (\up, \gp, \rp, \ip, \zp) to a limiting magnitude of r$<$22.5
\citep{1996AJ....111.1748F,1998AJ....116.3040G,2002AJ....123.2121S}.  Galaxy
clusters are selected from $\sim 7500$ sq. degrees of available SDSS
imaging data, and from the mock catalogs described below, using the
maxBCG method \citep{2007astro.ph..1268K} which is outlined in \S \ref{sec:maxbcg}.

The spectroscopic survey targets a
``main'' sample of galaxies with \rp$<$17.8 and a median redshift of
z$\sim$0.1 \citep{2002AJ....124.1810S} and a ``luminous red galaxy'' sample
\citep{2001AJ....122.2267E} which is roughly volume limited out to z=0.38, but
further extends to z=0.6.  The ``main'' sample composes about 90\%
of the catalog, with the ``luminous red galaxy'' sample making up the
rest. Velocity errors in the redshift survey are $\sim$30 \kms.  We
use the SDSS DR5 spectroscopic catalog which includes over 640,000
galaxies.  The mask for our spectroscopic catalog was taken from the
New York University Value-Added Galaxy Catalog \citep{2005AJ....129.2562B}.

\subsection{Mock Galaxy Catalogs} 
In order to understand the robustness of our methods for measuring the
mean and scatter of the relation between cluster velocity dispersion
and richness, we perform several tests on realistic mock galaxy
catalogs.  Because the maxBCG method relies on measurements of galaxy
positions, luminosities, and colors and their clustering, these
catalogs must reproduce these aspects of the SDSS data in some detail.

In this work, we use mock catalogs created by the ADDGALS (Adding
Density-Determined Galaxies to Lightcone Simulations) method
(described by Wechsler et al. 2007, in preparation) which is
specifically designed for this
purpose.  These catalogs populate a dark matter light-cone simulation with
galaxies using an observationally-motivated biasing scheme.  Galaxies
are inserted in these simulations at the locations of individual dark
matter particles, subject to several empirical constraints.  The
relation between dark matter particles of a given over-density (on a
mass scale of $\sim 10^{13} \mathrm{M}_{\odot}$) is
connected to the two point correlation function of these particles.   This connection
is used to assign subsets of particles to galaxies using a probability distribution
$P(\delta|L_r)$, chosen to reproduce the luminosity-dependent
correlation function of galaxies as measured in the SDSS by \citet{2005ApJ...630....1Z}.
The number of galaxies of a given brightness placed within the
simulations is determined from the measured SDSS $r$-band galaxy  luminosity function
\citep{2003ApJ...592..819B}.  We consider galaxies brighter than  0.4
\lstar, because it is these galaxies that are counted in the  maxBCG richness estimate.
Finally, colors are assigned to each galaxy by measuring their local
galaxy density in redshift space, and assigning to them the colors of a real SDSS galaxy
with similar luminosity and local density (see also
\citealt{2004ApJ...614..533T}).  The local density measure used is the
fifth nearest neighbor galaxy in a magnitude and redshift slice, and for
SDSS galaxies is taken from a volume-limited sample of the CMU-Pitt DR4
Value Added Catalog\footnote{Available at $\mathrm{www.sdss.org/dr4/products/value\_added}$.}. 

This method produces mock galaxy catalogs whose galaxies
reproduce several properties of the observed SDSS galaxies.  In
particular, they follow the empirical galaxy color--density relation and its
evolution, a property of fundamental importance for ridgeline-based
cluster detection methods.  The process accounts for $k$-corrections
between rest and observed frame colors and assigns realistic
photometric errors. Each mock galaxy is associated with a dark matter
particle and adopts its 3D motion. This is important, as it
encodes in the motions of the mock galaxies the full dynamical
richness of the N-body simulation. Galaxies may occupy fully
virialized regions, be descending into clusters for the first time, or
be slowly streaming along nearby filaments. This complete sampling of
the velocity field around fully realized N-body halos is essential, as
these mock catalogs allow us to predict directly the velocity
structure we ought to observe in the data. Note that this simulation
process, by design, assumes no velocity bias between the dark matter
and the luminous galaxies, except for the BCG, which is made by
artificially placing the brightest galaxy assigned to a given halo at
its dynamic center.

In this work, we use two mock catalogs based on different
simulations.  The first is based on the Hubble Volume simulation
\citep[the MS lightcone of][]{2002ApJ...573....7E}, which has a particle mass of
$2.25\times 10^{12}$\msun\, while the second is based on a simulation
run at Los Alamos National Laboratory (LANL) using the Hashed-Oct-Tree
code \citep{1994PhDT.........1W}.  This simulation tracks the evolution of
$384^3$ particles with $6.67\times10^{11}$\msun\ in a box of side
length 768 Mpc h$^{-1}$, and is referred to as the ``higher
resolution'' simulation in what follows.  Both simulations have
cosmological parameters $\Omega_m=0.3$, $\Omega_{\Lambda} = 0.7$, $h =
0.7$ and $\sigma_8 = 0.9$.

In addition to the galaxy list we have a list of dark matter halos,
defined using a spherical over-density cluster finder
\citep[e.g.][]{2002ApJ...573....7E}. By running the cluster finding algorithm on the
mock catalogs, we connect clusters detected
``observationally'' from their galaxy content with simulated dark
matter halos in a direct way.

\subsection{Velocity Bias}
Given that there is still substantial uncertainty in the amount of
velocity bias for various galaxy samples, we must be careful to avoid
velocity bias dependent conclusions.  The mock catalogs with which we
are comparing do not explicitly include velocity bias.  Fortunately,
we will only incur errors due to velocity bias when we estimate masses
or directly compare velocity dispersions measured in the mock catalogs
to those measured in the data.  Therefore in most of our analysis,
velocity bias has no effect.  Where it is relevant, we choose to leave
velocity bias as a free parameter because of its current observational
and theoretical uncertainties.

Observational uncertainties in velocity bias arise simply because it
is exceedingly difficult to measure.  To make such a measurement, one
usually requires two independent determinations of mass, one of them
based on dynamical measurements, each subject to systematic and random
errors \citep[e.g.][]{1994ApJ...434L..51C}.  Another
technique was recently used by \citet{2006astro.ph..6545R}, namely
constraining the
velocity bias by measuring the virial mass function and comparing it
to other independent cosmological constraints.  Their analysis resulted
in a bias of $b_{v}\sim1.1-1.3$.  Unfortunately, this technique folds
in systematic errors from the other analyses.

In the past, theoretical predictions of velocity bias were affected by
numerical over-merging and low resolution \citep[e.g.][]{1996ApJ...472..460F,2000ApJ...544..616G}.
Most estimates of velocity bias based on high-resolution N-body
simulations have given $b_{v}\sim1.0-1.3$
\citep{2000ApJ...539..561C,2000ApJ...544..616G,2004MNRAS.352..535D,2005MNRAS.358..139F},
partially depending on the mass regime studied.  Recent theoretical
work has shown that differing methods of subhalo selection in N-body
simulations change the derived velocity bias. In particular, \citet{2006MNRAS.369.1698F}
have shown that when subhalos are selected by their properties at the
time of accretion onto their hosts (a model which also matches the
two-point clustering better, see \citealt{2006ApJ...647..201C}), they are consistent
with being unbiased with respect to the dark matter.  Still,
understanding velocity bias with confidence will require more
observational and theoretical study.  As a result, we leave velocity
bias as a free parameter where assumptions are required.

\section{The MaxBCG Cluster Catalog}\label{sec:maxbcg}
\subsection{The maxBCG Cluster Detection Algorithm}

The maxBCG cluster detection algorithm identifies clusters as
significant over-densities in position-color space
\citep{2007astro.ph..1265K,2007astro.ph..1268K}. It relies on the fact
that massive clusters are dominated by bright, red, passively-evolving
ellipticals, known as the red-sequence \citep{2000AJ....120.2148G}. In
addition, it exploits the spatial clustering of red-sequence and the presence
of a cD-like brightest cluster galaxy (BCG). The brightest of the
red-sequence galaxies form a color--magnitude relation, the E/S0
ridgeline \citep{1999AAS...195.1202A}, whose color is a strong
function of redshift. Thus, in addition to reliably detecting
clusters, maxBCG also returns accurate photometric redshifts.  The
details of the algorithm can be found in \citet{2007astro.ph..1268K}.

The primary parameter returned by the maxBCG cluster detection
algorithm is \ngrtwo, the number of E/SO ridgeline galaxies dimmer
than the BCG, within +/- 0.02 in redshift (as estimated by the
algorithm), and within a scale radius $R_{200}^{\mathcal{N}}$ \citep{2005ApJ...633..122H}:
\begin{equation}
R_{200}^{\mathcal{N}} = (140\,h^{-1}kpc) \times N_{gal}^{0.55}
\end{equation}
where \ngal\ is the number of E/SO ridgeline galaxies dimmer than the
BCG, within +/- 0.02 in redshift, and within 1 $h^{-1}$Mpc.

The value of $R_{200}^{\mathcal{N}}$ is defined by \citet{2005ApJ...633..122H} as
the radius at which the galaxy number density of the cluster is
200$\Omega_{m}^{-1}$ times the mean galaxy space density.  This radius
may not be physically equivalent to the standard \rtwo\ defined as the
radius in which the total matter density of the cluster is 200 times
the critical density.  

In the work below, we also use the results of
Sheldon et al. (2007, in preparation) and Johnston et al. (2007, in
preparation) who calculate \rtwo\ from weak lensing analysis on
stacked maxBCG clusters.  \citet{2007ApJ...656...27J} show that these weak lensing
measurements can be non-parametrically inverted to obtain
three-dimensional, average mass profiles.  In the context of the halo
model, these mass profiles are fit with a one- and two-halo term.  The
best fit NFW profile \citep{1997ApJ...490..493N}, which comprises the 
one-halo term, is used to measure R200.  Several systematic errors are
accounted for including non-linear shear, cluster mis-centering, and
the contribution of the BCG light (modeled as a point mass).

It is notable that the redshift estimates for the maxBCG cluster
sample are quite good. They can be tested with SDSS data by comparing
them to spectroscopic redshifts for a large number of BCG galaxies
obtained as part of the SDSS itself.  The photometric redshift errors
are a function of cluster richness, varying from $\delta$z = 0.02 for
systems of a few galaxies to $\delta z \le 0.01$ for rich clusters
\citep{2007astro.ph..1265K}.

\citet{2007astro.ph..1265K,2007astro.ph..1268K} estimate the
completeness (fraction of real dark matter
halos identified) as a function of halo mass and purity (fraction of
clusters identified which are real dark matter halos) as a function of
\ngrtwo\ by running the detection algorithm defined above on the
ADDGALS simulations. The maxBCG cluster catalog is demonstrated
to have a completeness of greater than $90\%$ for dark matter halos
above a mass of $2 \times 10^{14}$ \msun, and a purity of greater than
90\% for detected clusters with observed richness greater than
\ngrtwo=10.  The selection function has been further characterized for use
in cosmological constraints by \cite{2007astro.ph..3574R}.

We finally note that clusters at lower redshift are more easily
identified directly from the spectroscopic sample (e.g. the C4
catalog, \citealt{2005AJ....130..968M} or the catalog of
\citealt{2006ApJS..167....1B}), but
are limited in number due to the high flux limit of the spectroscopic
sample. Clusters at redshifts higher than 0.3 can be identified easily
in SDSS photometric data, but measurement of their richnesses,
locations, and redshifts {\it in a uniform way} becomes increasingly
difficult as their member galaxies become faint. Future studies
similar to the one described in this paper will be possible as the
maxBCG method is pushed to higher redshift and higher redshift
spectroscopy is obtained.

\subsection{The Cluster Catalog}
The published catalog \citep{2007astro.ph..1265K} includes a total of 13,823
clusters from $\sim7500$ square degrees of the SDSS, with $0.1 < z <
0.3$ and richnesses greater than \ngrtwo=10. For
this study, we extend the range of this catalog to $0.05 < z < 0.31$
and \ngrtwo$ \geq 3$.  The lower redshift bound allows us to
include more of the SDSS spectroscopy, which peaks in density around
$z \sim 0.1$.  The extended catalog used in this study sacrifices the well-understood
selection function of the maxBCG clusters for the extra spectroscopic
coverage and thus improved statistics.  The lower richness cut additionally
cut allows us to probe a wider range of cluster and group masses. This
larger sample has a total of 195,414 clusters and groups.

The selection function has only been very well characterized (by
\citealt{2007astro.ph..1265K,2007astro.ph..1268K} and
\citealt{2007astro.ph..3574R}) for the maxBCG catalog presented
in \citet{2007astro.ph..1265K}. The broader redshift range and lower richness limit
considered for this study are not encompassed in the preceding
studies.  This is primarily because we expect the color selection may produce less
complete samples for low richness; since the red fraction in clusters
and groups decreases with decreasing mass, maxBCG may be biased against
the bluest low mass groups.

Requiring sufficient spectroscopic coverage for each cluster, defined in \S
\ref{sec:cgcf} in the context of the construction of the BCG--galaxy
velocity correlation function, significantly restricts the sample of
clusters studied here due to the
limited spectroscopic coverage of the SDSS in comparison with its
photometric coverage.  Most of the maxBCG clusters above $z\approx0.2$ contribute
relatively little to the \cgcf.  The final cluster sample includes
only 12,253 clusters. A total of 57,298 of the more than 640,000 SDSS
DR5 galaxy redshifts are used in this study.

\section{The Velocity dispersion--Richness relation}\label{sec:observables}
To compare cluster catalogs derived from data to theoretical
predictions of the cluster mass function, we must examine cluster
observables which are related to mass. For individual clusters, the
primary mass indicators we have for this photometrically-selected
catalog are based on observations of galaxy content.  Some of the
observable parameters include \ngal, total optical luminosity \lo, and
comparable parameters measured within observationally scaled radii
\ngrtwo\ and \lortwo. To understand the relationship between these
various richness measures and cluster mass, we can refer to several
observables more directly connected to mass: the dynamics of galaxies,
X-ray emission, and weak lensing distortions
the clusters produce in the images of background galaxies. In this
work we concentrate on the extraction of dynamical information from
the maxBCG cluster catalog. Weak lensing
measurements of this cluster catalog are described by Sheldon et
al. (2007, in preparation) and Johnston et al. (2007, in preparation).
An analysis of the average X-ray
emission by maxBCG clusters is in preparation (Rykoff et al. 2007).
Preliminary cosmological constraints from this catalog, based only on cluster counts,
have been presented by \citet{2007astro.ph..3571R}; these will be extended with the additon
of these various mass estimators.

\subsection{Extracting Dynamical Information from Clusters: the 
BCG--Galaxy Velocity Correlation Function}
\label{sec:cgcf}
Using the SDSS spectroscopic catalog, we can learn about the dynamics
of the maxBCG galaxy clusters. For this sample, drawn from a redshift
range from 0.05 to 0.31, the spectroscopic coverage of cluster members
is generally too sparse to allow for direct measurement of individual
cluster velocity dispersions.  We instead focus here
on the measurement of the mean motions of galaxies as a function of
cluster richness. We study these motions by first constructing the
BCG--galaxy velocity correlation function, $\xi(\delta v, r,
P_{cl}, P_{gal})$, hereafter, the BCG--galaxy velocity correlation function, BGVCF.

To construct the \cgcf, we identify those clusters for which a BCG
spectroscopic redshift has been measured.  We then search for
other galaxies with spectroscopic redshifts contained within a
cylinder in redshift-projected separation space which is $\pm 7,000$
\kms deep and has a radius of one \rtwo, which varies as a function of
\ngrtwo, as measured by Johnston et al. (2007, in preparation). For each such spectroscopic
neighbor we form a ``pair'', recording the velocity separation of the
pair, $\delta v$, their
projected separation at the BCG redshift, $r$, information about the
properties of each galaxy (the BCG and its neighbor)
$P_{gal}$, and information about the cluster in which the BCG resides
$P_{cl}$.  This pair structure contains the observational
information relevant to the \cgcf.

\begin{figure}
  \plotone{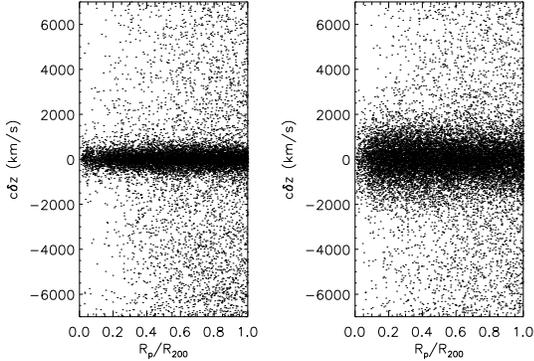} \figcaption[f1.eps]{The projected separation
    and velocity separation for the pairs of galaxies in clusters with
    \ngrtwo\ $\geq$ 15 (right) and \ngrtwo\ $\leq$ 5 (left). There is
    a clear change in the \cgcf\ with \ngrtwo.\label{fig:rdgrms}}
\end{figure}

The quantities $P_{gal}$ and $P_{cl}$ will change depending on the
context in which we are considering the \cgcf.  Some examples of
$P_{cl}$ include \ngal, \lo, \ngrtwo, \lortwo, local environmental
density, and \rtwo.  Examples of $P_{gal}$ include the magnitude
differences between members and the BCG, BCG $i$-band luminosity, and
stellar velocity dispersion.  The mean of $\delta v$ is consistent
with zero so that the \cgcf\ is independent of the parity of $\delta
v$.  In Figure \ref{fig:rdgrms} we show the \cgcf\ of the catalog in
two bins of \ngrtwo, one with \ngrtwo\ $< 5$ (left
panel) and one with \ngrtwo\ $> 15$ (right panel).  The structure of the \cgcf\
clearly changes with richness.
 
\begin{figure}[b]
\plotone{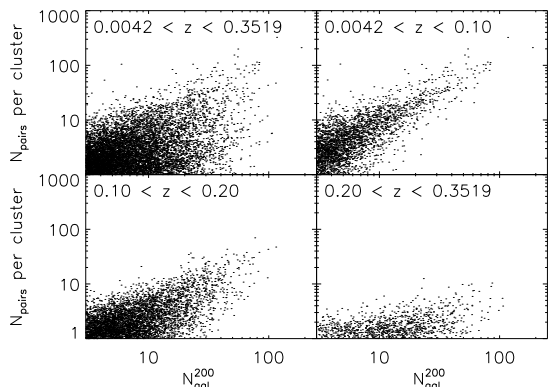}
\figcaption[f2.eps]{The number of spectroscopic pairs in the
  \cgcf\ per cluster as a function of redshift.  The redshift ranges
  for each panel are indicated; the first panel includes the
  entire catalog.  Clusters with lower redshifts and higher richness
  are better sampled by the spectroscopic survey.   The points in this
  diagram are displaced randomly from their integral values
  (i.e. $\{1, 2, ...\}$) so that
  the true density can be seen.\label{fig:npairsred}} 
\end{figure}

When we stack clusters to measure their velocity dispersion as
described below, the statistical properties of our sampling of the \cgcf\
determine the errors in our measurements.  Figures
\ref{fig:npairsred}(a)-(d) show the number of pairs in the
\cgcf\ per cluster as a function of \ngrtwo\ plotted for the entire
\cgcf\ and in three redshift bins.  As the redshift of the bins
increases, we can see that the number of spectroscopic pairs becomes
less reflective of the value of \ngrtwo\ for the cluster.  Clusters at
lower redshift tend to have more pairs, as expected.  Figure
\ref{fig:npairsred} shows that if we want to measure the velocity
dispersion of individual clusters, we are limited to low redshift and
high richness because only these clusters are sufficiently well
sampled by the SDSS spectroscopic data. 
 
\subsection{Characterizing the BGVCF of Stacked Clusters}
\label{sec:cgcfchar}
In this work, we are primarily concerned with the magnitude of the velocity
dispersion and its scatter at fixed richness, as well as its dependence on the properties of
clusters and their galaxies.  To greatly simplify our analysis, we now
integrate the \cgcf\ radially to produce the pairwise velocity
difference histogram (PVD histogram).  Strictly speaking, we do not
produce a true PVD histogram because the only pairs we consider are
those between BCGs and non-BCGs around the same cluster (i.e. all
other galaxies in the \cgcf\ around each cluster).  We do not include
non-BCG to non-BCG pairs.

Ideally, if every cluster had a properly selected BCG and all BCGs
were at rest with respect to the center-of-mass of the cluster, our
measurements of the mean velocity dispersion would be unbiased with
respect to the true center-of-mass velocity dispersion.  Unfortunately,
these simplifying assumptions are not likely to be true.
In particular it has been found that
BCGs move on average with $\sim25\%$ of the cluster's velocity
dispersion, but that at higher mass BCG movement becomes more
significant \citep[e.g.][]{2005MNRAS.361.1203V}.  In
$\S$\ref{sec:bias} we show that a correction must be
applied to our mean velocity dispersions due to centering on the BCG
(hereafter called BCG bias), but that we cannot
distinguish between improperly selected BCGs and BCG movement.
However, we will still focus on the BCG in the measurements of the
\cgcf\ because it is a natural center for the cluster in the context
of the maxBCG cluster detection algorithm.

Having decided to concentrate on the PVD histogram, we now motivate
the construction of a fitting algorithm for the PVD histogram.
Previous work by \citet{2002ApJ...571L..85M}, \citet{2003ApJ...598..260P}, and others
\citep{2003ApJ...593L...7B,2004MNRAS.352.1302V,2005ApJ...635..982C,
 2007ApJ...654..153C} has focused on measuring the
halo mass of isolated galaxies by using dynamical measurements.
\citet{2002ApJ...571L..85M} found the velocity dispersion around these galaxies by
stacking them in luminosity bins and fitting a Gaussian curve plus a
constant, representing the constant interloper background, to the
stacked PVD histogram.  In this method, the standard deviation of the
fit Gaussian curve is then taken as an estimate of the mean value of the
velocity dispersion of the stacked groups.

\begin{figure*}
\plotone{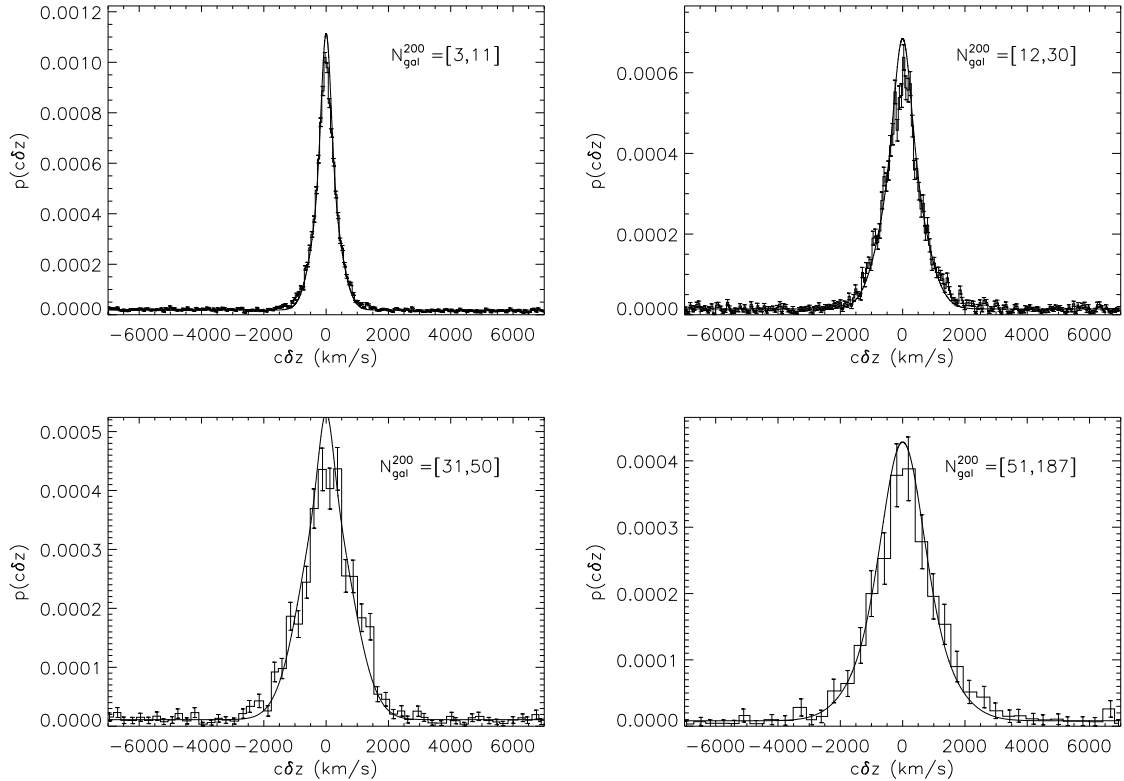}
\figcaption[f3.eps]{PVD histograms in four \ngrtwo\ bins.  The EM
  algorithm fits are shown along with the Poisson errors of the
  histograms.  Notice the change in the degree of non-Gaussianity of
  the lower \ngrtwo\ bins compared to the higher \ngrtwo\ bins.  As
  richness increases, the stacked PVD histogram becomes more Gaussian.
  Section \ref{sec:nongauss} shows that this decrease indicates a
  decrease the width of the lognormal distribution of velocity
  dispersions in each bin.  The deviations of the fits near the
  centers of the distributions are only at the two-sigma level and are
  highly dependent on the bin size used to produce the
  cluster-weighted PVD histograms.\label{fig:pvd4b4}}
\end{figure*}

The algorithm presented above is insufficient for our purposes for the
following reason.  The PVD histogram
of stacked galaxy clusters is shown in Figure \ref{fig:pvd4b4}; it is
clearly non-Gaussian.  Although the width of a single Gaussian
curve likely still provides some information about the typical
dispersion of the sample, it cannot adequately capture the
information contained in the non-Gaussian shape of the PVD histogram.
As we will show below in \S \ref{sec:nongauss}, although the PVD
histogram for a stack of similar velocity dispersion clusters is
expected to be nearly Gaussian, there are multiple sources of
non-Gaussianity that can contribute to the non-Gaussian shape of the
stacked PVD histograms.  To adequately characterize this non-Gaussianity, one
of the primary goals of this paper, we
must use a better fitting algorithm to characterize the PVD
histograms.
 
In this work we will mention a variety of different methods of
fitting the PVD histograms.  We
give their names and definitions here and follow with a full description
of the primary method used, 2GAUSS.  The various methods are:
\begin{description}
\item[1GAUSS:] This is the method used for isolated galaxies as
  discussed above.  We do not use it because, as described in \S
  \ref{sec:fittest}, it systematically underestimates the second
  moment of the PVD histogram by $\sim8\%$.
\item[2GAUSS:] This method is the one motivated and described in
  detail below.  It is the primary method used throughout the rest of
  the paper.  Simply, it fits the PVD histogram with two Gaussians and
  a constant background term, but with a special weighting by cluster
  and not by galaxy (see \S \ref{sec:nongauss}).
\item[NGAUSS:] This is a generalized version of the 2GAUSS method with
  N Gaussians instead of two (i.e. a three Gaussian fit will be
  denoted by 3GAUSS).  Although it fits the PVD histogram as well as
  the 2GAUSS method, it is more computationally expensive, and adds
  parameter degeneracies without substantially improving the quality of
  the fit.
\item[NONPAR:] There are several possible methods for using
  non-parametric fits to the PVD histogram (e.g. kernel density
  estimators).  We do not use them because they do not naturally
  account for the constant interloper background in the PVD histogram.
  For a good review of these techniques see \citet{2001astro.ph.12050W}.
\item[BISIGMA:] This method is not used for the PVD histograms of
  stacked clusters, but is used for the PVD histograms of individual
  clusters.  The biweight is a robust estimator of the standard
  deviation that is appropriate for use with samples of points which
  contain interlopers \citep{1990AJ....100...32B}.  See \S \ref{sec:icvds} for a
  description of its use in this paper.
\item[BAYMIX:] This method is a Bayesian or maximum likelihood method
  that can be used in the context of the model of the scatter in
  velocity dispersion at fixed richness.  This method will be
  described fully in \S \ref{sec:nongauss}, but we do not use it in
  this paper because we have found it to be unstable.
\end{description}
We will refer to these methods by their names given above.  Although
we mention these other methods, for deriving the main results of the
paper we use the 2GAUSS method for stacked cluster samples
(\ref{sec:twogauss}) and the BISIGMA method for individual clusters (\S
\ref{sec:icvds}).

\subsubsection{Fitting the PVD Histogram}
\label{sec:twogauss}
In order to more fully capture the shape of the PVD histogram of
stacked clusters, avoid systematic fitting errors, and avoid fitting
degeneracies, we would ideally use the NONPAR method to fit the PVD
histogram.  In this way we would impose no particular form on
the PVD histogram, allowing us to extract its true shape with as few
assumptions as possible.  However, this method does not naturally
account for the interloper background term of the PVD histogram which
can be easily fit by a constant \citep{2006astro.ph..6579W}.   

In the pursuit of simplicity, we compromise by fitting the PVD
histogram of stacked clusters with two Gaussian curves plus a constant
background term.  The means of the two Gaussians are free
parameters but are fixed to be equal.  In all cases the
mean is consistent with zero.  The two Gaussian curves allow us to
more fully capture the shape of the PVD histogram, while still
accounting for the interloper background of the \cgcf\ with the
constant term.  It could be that the shape of the PVD histogram cannot
be satisfactorily accounted for by two Gaussians.  We show in \S
\ref{sec:fittest} that two Gaussians are sufficient to describe the
shape of PVD histogram.  Using the 2GAUSS method instead of
the NGAUSS method avoids expensive computations and limits the number
of parameters in the fitting procedure to six, avoiding degeneracies
in the fit parameters due to limited statistics.

In the interest of fitting stability and
ease of use (but sacrificing speed), we use the expectation
maximization algorithm (EM) for one dimensional Gaussian mixtures to fit
the PVD histogram \citep{dm77,2000astro.ph..8187C}.  In Appendix
\ref{app:emalg}, we re-derive the EM algorithm for one dimensional
Gaussian mixtures such that it assigns every Gaussian the same mean
and weights groups of galaxies, not individual galaxies, evenly.  This last step
is important in the context of the model of the distribution of
velocity dispersions at fixed \ngrtwo\ discussed in \S
\ref{sec:nongauss}.  To account for our velocity errors, we use the
results of \citet{2000astro.ph..8187C} and subtract in quadrature the $30$ \kms
redshift error from the standard deviation of each fit Gaussian. 

We have described our measurements in the context of the PVD histogram
and not the \cgcf.  However, these two view points in our case are
completely equivalent.  \citet{2006astro.ph..6579W} have shown that galaxies
uncorrelated with the cluster in PVD histograms (i.e. interlopers) form a constant
background.  Thus by fitting a constant term to the PVD histogram, we
are in effect subtracting out the uncorrelated pairs statistically to
retain the \cgcf.

\subsubsection{Tests of the 2GAUSS Fitting Algorithm}\label{sec:fittest}
In order to measure the moments of the PVD histogram as
a function of richness, the data is first binned logarithmically in
\ngrtwo\ and then the 2GAUSS method is applied to each bin.  The
results of our fitting on four bins of \ngrtwo\ are shown in Figure
\ref{fig:pvd4b4}.  In all cases, the model provides a reasonable fit
to the data.  We defer a full discussion of the fitting results to \S
\ref{sec:nongauss} where we show how to compute the mean velocity
dispersion and scatter in velocity dispersion at fixed \ngrtwo\ using
the results of the 2GAUSS fitting algorithm, including corrections for
improperly selected BCG centers and/or BCG movement.

To ensure that the use of the 2GAUSS method does not bias our fits
in any way, we repeated them using the 1GAUSS, 3GAUSS, and 4GAUSS
methods.  We find that while the measured second moment for the 1GAUSS
fits are consistently lower than those measured from the 2GAUSS fits
by approximately 8\%, both the second and fourth moments measured by
the 2GAUSS, 3GAUSS, and 4GAUSS fits are the same to within a few
percent.  Therefore we conclude that the fits have converged and that
two Gaussians plus a constant are sufficient to capture the overall
shape of the PVD histogram.  The fitting errors are determined using bootstrap
resampling over the clusters in each bin.

The results are not dependent on the radial or velocity scale used to
construct the \cgcf\ and thus the PVD histogram.  We repeated the
2GAUSS fits using 0.75\rtwo, 1.0\rtwo, 1.25\rtwo, and 1.5\rtwo\
projected radial cuts as well as $\pm10000$ \kms, $\pm5\sigma$ scaled,
and $\pm10\sigma$ scaled apertures in velocity space.  We found no
significant differences in the fits using each of the various cuts,
with the exception of the value of the background normalization,
which will change when a larger aperture allows more background to be
included in the PVD histogram.  The scaled apertures were made by
first determining the relation in a fixed aperture, and then rescaling
the aperture according to this relation.  For example, in a bin of
\ngrtwo\ from 18 to 20, the velocity dispersion is $\sim500$ \kms as
measured in a $\pm$7000 \kms fixed aperture.  To make the five sigma
scaled aperture measurements, we used an aperture in this \ngrtwo\ bin
of $\pm5\times500=\pm2500$ \kms.

\subsection{Estimating Individual Cluster Velocity Dispersions}\label{sec:icvds}
\begin{figure}
\plotone{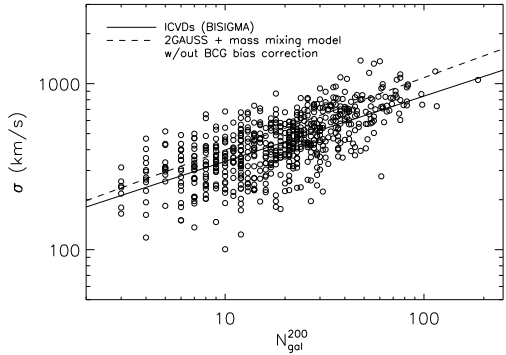}
\figcaption[f4.eps]{The individual cluster
  velocity dispersions for the entire data set.  The dispersions were
  computed using the BISIGMA method on only those clusters with more
  than ten \cgcf\ pairs within three-sigma of the BCG
  (calculated using the mean velocity dispersion at the BCG's
   \ngrtwo).  The dashed line is the geometric mean of the velocity
  dispersions calculated in the context of the mixing model (\S
  \ref{sec:nongauss}) without a correction for BCG bias and the solid
  line is the fit to all velocity dispersions of the individual
  clusters.  The error bars on the ICVDs are not shown for clarity.
  The two relations (dashed and solid) agree with each other within
  one- to two-sigma, but the average bias between them is a real
  effect.\label{fig:icvds}}  
\end{figure}

To measure the velocity dispersion of individual clusters in the SDSS,
we select all clusters that have at least ten redshifts in its PVD
histogram within three sigma measured by the mean
velocity-dispersion--\ngrtwo\ relation calculated in \S
\ref{sec:nongauss}.  Of the 12,253 clusters represented in the \cgcf,
only 634 meet the above requirement. We then apply the BISIGMA method to calculate the
velocity dispersion which uses the biweight estimator
\citep{1990AJ....100...32B}.  The resulting velocity dispersions are
plotted in Figure \ref{fig:icvds}.  The BCG bias manifests itself here
in that the ICVDs show a downward bias with respect to the mean
relation calculated in \S \ref{sec:nongauss}, but not corrected for
BCG bias.  We will correct for this bias in \S \ref{sec:bias}.  The
two relations do however agree to within one- to
two-sigma (computed through jackknife resampling with the biweight).
Using these individual cluster velocity dispersions (ICVDs), we can
directly compute the scatter in the velocity dispersion at fixed
\ngrtwo.  We will compare this computation with the estimate based on
measuring non-Gaussianity in the stacked sample in \S \ref{sec:nongauss}.
 
\section{Measuring Scatter in the Velocity Dispersion--Richness Relation}
\label{sec:nongauss}

\subsection{Mass Mixing Model}
The subject of non-Gaussianity of pair-wise velocity difference
histograms has been debated extensively in the literature.
\citet{1996ApJ...467...19D} have shown that non-Gaussianity in the total PVD
histogram for dark matter halos arises from two sources: stacking
halos of different masses according to the mass function, and
intrinsic non-Gaussianity in the PVD histogram due to substructure,
secondary infall, and dissipation of orbital kinetic energy into
subhalo internal degrees of freedom. However for galaxy clusters, they
conclude that the PVD histogram of an individual galaxy cluster that
is virialized is well approximated by a Gaussian.

\citet{1996MNRAS.279.1310S} independently reached the same conclusions but did not
consider any intrinsic non-Gaussianity, just the effect of stacking
halos of different masses according to the mass function.
\citet{2001MNRAS.322..901S} gave a more complete synthesis of
non-Gaussianity in PVD histograms, generalizing the formalism to
include the effect of particle tracer type (halos versus galaxies
versus dark matter particles) and extensively considering the effects
of local environment.  In all three treatments, non-Gaussianity is
shown to arise from the stacking of individual PVD histograms which
are Gaussian or nearly Gaussian and have some intrinsic distribution
of widths.  In this paper we use the term ``mass mixing'' to refer
only to non-Gaussianity arising through this process.

It should be noted that observing non-Gaussianity in PVD histograms
stacked by richness is equivalent to saying that the richness verses
velocity dispersion relation has intrinsic scatter (assuming that the
stacked PVD histogram of set of similar velocity dispersion clusters
has intrinsic Gaussianity).  Intrinsic scatter in the
velocity-dispersion--richness relation was reported earlier by
\citet{1996A&A...310...31M} for a volume-limited sample of 80
literature-selected clusters with at least 10 redshifts each.  Here we
seek to quantify this scatter as a function of richness by measuring
the non-Gaussianity in the PVD histogram.

For individual galaxy clusters, theoretical work has been done by
\citet{2005PhRvE..71a6102I} showing that violent gravitational collapse in an N-body
system may lead to a non-Gaussian velocity distribution.  However,
\citet{2006MNRAS.369.1698F} have shown that the velocity distribution of subhalos in
a dissipationless N-body simulation is Gaussian (Maxwellian in three
dimensions) and shows little bias compared to the diffuse dark matter,
if the subhalos are selected by their mass when they enter the host
halo and not the present-day mass.  Finally,
\citet{2001MNRAS.322..901S} caution against concluding that the
three-dimensional velocity distribution of a galaxy cluster is
Maxwellian even if one component is found to be approximately
Gaussian.  They show that for a slightly
non-Maxwellian three-dimensional distribution, departures of the
one-dimensional distribution from a Gaussian are much smaller than
departures of the three-dimensional distribution from a Maxwellian. 

Observationally, the PVD histogram of individual galaxy clusters has
been shown to be non-Gaussian in the presence of substructure
\citep[e.g.][]{2004A&A...425..429C,2004A&A...427..397H,
  2005A&A...442...29G}.  Conversely, \citet{1993ApJ...404...38G}
observed 79 galaxy clusters with at least 30 redshifts each and found
no systematic deviations from Gaussianity (although 14 were found to
be mildly non-Gaussian at the three-sigma level).
\citet{1987AJ.....94.1116C} showed that once recently-accreted
galaxies are removed from the sample of redshifts from the Fornax
cluster, the PVD histogram becomes Gaussian. 

Unfortunately, due to the low number of galaxy redshifts available for
a given cluster as shown in \S \ref{sec:cgcf}, we must make some
assumption about the shape of the PVD histogram for \textit{a set of
  stacked clusters of velocity dispersion between $\sigma$ and $\sigma
  + d\sigma$} in order to proceed
with constructing a mass mixing model.  If every cluster were sampled
sufficiently, the scatter in velocity dispersion at a given value of
\ngrtwo\ could be directly computed by measuring the velocity
dispersions of individual clusters.  Since this is not the case, in
order to proceed we assume that for \textit{a set of stacked clusters
  of velocity dispersion between $\sigma$ and $\sigma + d\sigma$}, the
PVD histogram is Gaussian.  This assumption is well justified and is
equivalent to the assumption that a large enough portion of the
clusters in our catalog are sufficiently relaxed, virialized systems
at their centers, so that when we stack them, any asymmetries or
substructure are averaged out.

However, we will still be sensitive to substructure around the BCG.
In fact, we may even be more sensitive to substructure around the BCG
since we are directly stacking clusters on the BCG.  Using the
ADDGALS mock galaxy catalogs, we find that when binning dark
matter halos with galaxies by both velocity dispersion and mass, the
resulting stacked PVD histograms are Gaussian.  This result gives us
further confidence that the above assumption is reasonable, but it is
still possible that it is sensitive either to the BCG placement or to
the galaxy selection of the mock catalogs.

Under the assumption of Gaussianity of the PVD histogram for a stacked
set of similar velocity dispersion clusters, the non-Gaussianity in
the stacked histograms can be entirely attributed to the distribution
of the velocity dispersions (or equivalently mass) of the stacked
clusters.  The goal of our analysis is then to extract the
distribution of velocity dispersions for a given PVD histogram by
measuring its deviation from Gaussianity.

We can now proceed in two distinct ways.  First, by writing the PVD
histogram as a convolution of a Gaussian curve of width $\sigma$ for
each stacked set of similar velocity dispersion clusters, with some
distribution of velocity dispersions in the stack, we could
numerically deconvolve this Gaussian out of the PVD histogram to
produce the distribution of velocity dispersions in the stack.
Repeating this procedure in various bins on different observables, we
could then have knowledge of the scatter in velocity dispersion as a
function of these observables. 

Second, by taking a more model dependent approach, we could make an
educated guess of the distribution of $\sigma$ as a function of a
given set of parameters, and then perform the convolution to predict
the shape of the PVD histogram.  By matching the predictions with the
observations through adjusting the set of parameters, we could then have a
parameterized model of the entire distribution.  

Based on the results shown below in \S \ref{sec:tests}, it
is apparent that the only parameter upon which $\sigma$ varies
significantly, neglecting the modest redshift dependence, is \ngrtwo.  Thus
we choose a parametric model that is a function of \ngrtwo\ only.  The
dependence of mass mixing on any secondary parameters (e.g. the BCG
$i$-band luminosity) can then be explored through first binning on
\ngrtwo\ and then splitting on these secondary parameters, because
their effects are small (see \S \ref{sec:tests}).

The ADDGALS mock catalogs show the scatter about the mean of
the logarithm of the velocity dispersion measured from the dark matter
for a given value of \ngal\ to be approximately Gaussian for dark
matter halos.  Using this distribution as our educated guess, we apply
this model to the data in logarithmic bins of \ngrtwo.  We avoid
the deconvolution due to its inherent numerical difficulty.  Using the
mock catalogs, we can test our method of determining the parameters of
this model as a function of \ngrtwo\ by
applying our analysis to clusters identified in the mocks and then
matching those clusters to halos in order to determine their true
velocity dispersions.

To summarize, there are two and possibly even three sources of non-Gaussianity in
our PVD histograms (similar to those discussed in \citet{1996MNRAS.279.1310S},
\citet{1996ApJ...467...19D}, and \citet{2001MNRAS.322..901S} discussed
above): (1) intrinsic non-Gaussianity in the PVD histogram for an
individual galaxy cluster, (2) the range of velocity dispersions that
contribute to the PVD histogram for a given value of \ngrtwo\ (mass mixing), and (3)
stacking of clusters with different values of \ngrtwo\ in the same PVD
histogram.  We handle the last two sources of non-Gaussianity jointly
through the model below and ignore the first, which is expected to be
small, both because most clusters are relaxed, virialized systems and
many are stacked together here.

As a final note, by binning logarithmically in \ngrtwo\ and then
measuring the mass mixing, we only approximately account for
non-Gaussianity arising from clusters with different richnesses in the
same bin.  By avoiding binning all together and finding the model
parameters through a maximum likelihood approach, we could remedy this
issue.  This approach is the BAYMIX method.  However, we have found
this process to be computationally expensive and slightly unstable due
the integral in equation \ref{eqn:badint} below.  Its only true advantage is
in the computation of the errors in the parameters and their
covariances.  Using a maximum likelihood approach, one could calculate
the full covariance matrix of the parameters introduced below.  As
will be shown below, binning in \ngrtwo\ and then measuring the mass
mixing will allow us to only easily find the covariance matrices of the
parameters in sets of two.  Since we are not significantly concerned
with the exact form of these errors or the covariances of the
parameters, we choose to bin for simplicity.

Future analysis of this sort with PVD histograms will hopefully take a
less model-dependent approach by deconvolving a Gaussian directly from
the stacked PVD histogram.  This will allow for a direct confirmation
of the distribution of velocity dispersions at fixed richness.  Also,
a direct deconvolution would allow one to make mass mixing
measurements of stacks of clusters binned on any observable.  Although
the lognormal form assumed here may in fact have wider applicability,
we can only confirm its use for clusters stacked by richness.

\subsection{Results Using The Mass Mixing Model}
We can write the shape of the non-background part
of the stacked cluster-weighted PVD histogram, $P(v)$, as
\begin{equation}
P(v)=\int{p(v,\sigma)d\sigma}=\int{p(v|\sigma)p(\sigma)d\sigma} 
\end{equation}
where $v$ is the velocity separation value and $\sigma$ is the
Gaussian width of a stacked set of similar velocity dispersion clusters.
Using the assumptions from the previous discussion, we let
$p(v|\sigma)$ be a Gaussian of width $\sigma$ with mean zero, and
$p(\sigma)$ be a lognormal distribution.  Performing the convolution,
we get that $P(v)$ is given by
\begin{eqnarray}\label{eqn:badint}
\lefteqn{P(v)=}\nonumber\\
&&\int_{0}^{\infty}{\frac{1}{\sigma^{2}2\pi
    S}\exp{\left(-\frac{v^{2}}{2\sigma^{2}}-\frac{(\ln{\sigma}-<\!\ln{\sigma}\!>)^{2}}
    {2S^{2}}\right)}\,d\sigma}
\end{eqnarray}
where $<\!\ln{\sigma}\!>$ is the geometric mean of \sigv\ and $S$ is
the standard deviation of $\ln{\sigma}$.  We note that the quantity
$100\times S$ is the percent scatter in $\sigma$.  The second and
fourth moments of this PVD distribution, $\mu_{(2)}$ and $\mu_{(4)}$ are given by
\begin{equation}\label{eqn:mu2disp}
\mu_{(2)}=\exp{(2<\!\ln{\sigma}\!> + 2S^{2})}
\end{equation}
and
\begin{equation}
\mu_{(4)}=3\,\exp{(4<\!\ln{\sigma}\!> + 8S^{2})}\ .
\end{equation}
For convenience we define the normalized kurtosis to be
\begin{equation}\label{eqn:normkurt}
\gamma_{N}^{2}=\frac{\mu_{(4)}}{3\mu_{(2)}^{2}}=\exp{4S^{2}}\ .
\end{equation}
Note that the odd moments of this distribution are expected to vanish,
and in fact the data is consistent with both the first and third
moments being zero.  Equation \ref{eqn:mu2disp} shows us why the
velocity dispersions derived directly from the second moment of the
PVD histogram must be corrected.  The factor of $\exp{S^{2}}$
artificially increases the velocity dispersions.  In practice this
effect is at most $\sim20\%$ at low richness and declines to $\sim5\%$
for the most massive clusters in our sample.

To complete our model, we need a term corresponding to the background
of the PVD histogram.  This background has two parts, an uncorrelated
interloper component and an infall component (i.e. galaxies which are
not in virial equilibrium but are bound to the cluster in the infall
region).  \citet{2006astro.ph..6579W} have shown that the uncorrelated
background is a constant in the PVD histogram, while
\citet{2004MNRAS.352.1302V} have shown that the infall component is
not constant and forms a wider width component for PVD histograms
around isolated galaxies.  We ignore the possible infall components in
our PVD histograms but note that they may bias the widths of our
lognormal distributions high.  Investigation of this issue in the mock
catalogs shows that the result of \citet{2004MNRAS.352.1302V} holds
for galaxy clusters as well.  Although we do not explore this here, it
may be possible to reduce the mass mixing signal from infalling
galaxies by selecting galaxies by color (i.e. red galaxies only or
just maxBCG cluster members), which preferentially selects
galaxies near the centers of the clusters. 

Accounting for the constant interloper background, the full model of
the cluster-weighted PVD histogram, $\mathcal{P}(v)$, can now be
written as,
\begin{equation}
\label{eqn:model}
\mathcal{P}(v)=\frac{p}{2L} + (1-p)P(v)
\end{equation}
where $L$ is the maximum allowed separation in velocity between the
BCG and the cluster members, set to $7000$ \kms, and $p$ is a
weighting factor that sets the background level in the PVD histogram.
Here we ignore the small error in the normalization due to integrating
$\mathcal{P}(v)$ over $v$ from $-\infty$ to $\infty$ instead of $-L$
to $L$.  As long as $L$ is sufficiently large, say on the order of
$4\exp(<\!\ln{\sigma}\!>)$ for a given PVD histogram, this error is
small.

Now, it can be seen why the stacked PVD histogram must be weighted by
cluster instead of by galaxy.  In equation \ref{eqn:badint}, equal
weight is assigned to each cluster because $p(v|\sigma)$ is a Gaussian
normalized to integrate to unity.  In order to predict the
pair-weighted PVD histogram correctly, we would have to predict the
total number of \cgcf\ pairs, both cluster and background, as function
of \ngrtwo\ and include this total in the integral and the background
term.  (The factors of $p$ and $1-p$ take care of the relative
weighting of the background relative to cluster, assuming this
weighting is the same for every cluster.  This may not be true, in
which case the factors of $p$ and $1-p$ would have to be included in
the integral as well.)  This is a significant problem due to its
dependence on redshift, local environment, and the selection function
of the survey. 

By using the cluster-weighted PVD histogram fit by the EM algorithm
derived in Appendix \ref{app:emalg} (i.e. the 2GAUSS method), we can
avoid this issue.  This weighting could be included in the BAYMIX
method as well. We choose to use the 2GAUSS method because it is more
stable and less computationally expensive.  In practice the extra
weighting factors do not change our results drastically, indicating
that most clusters already get approximately equivalent weight even in
the pair-weighted PVD histogram.  However, for completeness we include
the weighting factor.  Note that two Gaussians is the the fewest
number of Gaussians a distribution could be composed of and have a
normalized kurtosis different from unity (the normalized kurtosis of a
single Gaussian distribution is unity).  According to equation
\ref{eqn:normkurt}, then, if we measure a normalized kurtosis of unity
for any of our bins, mass mixing in that bin (i.e. $S$) will be zero.

The quantities $<\!\ln{\sigma}\!>$, $S^{2}$, and $p$ are measured by
binning the data in \ngrtwo\ and applying the 2GAUSS method as
described earlier.  This method outputs the constant background level
$p$ automatically.  The normalized kurtosis is calculated as
\begin{equation}\label{eqn:kurttngal}
\gamma_{mes}^{2}=\frac{p_{1}(\sigma_{1})^{4} +
  p_{2}(\sigma_{2})^{4}}{\left(p_{1}(\sigma_{1})^{2} +
  p_{2}(\sigma_{2})^{2}\right)^{2}}
\end{equation}
and the second moment is calculated as
\begin{equation}\label{eqn:sigmatngal}
\mu_{(2)}^{mes}=\frac{p_{1}(\sigma_{1})^{2} +
  p_{2}(\sigma_{2})^{2}}{p_{1}+p_{2}}\ ,
\end{equation}
where $\{p_1,p_2\}$ and $\{\sigma_1,\sigma_2\}$ are the normalizations
and standard deviations calculated for the two Gaussians in the 2GAUSS
PVD histogram fit.  See Appendix \ref{app:emalg} for more
details. Although equation \ref{eqn:kurttngal} is not properly
normalized, we find that the bias correction computed in Appendix
\ref{app:bias} is small, and thus equation \ref{eqn:kurttngal} is a
good estimator of the normalized kurtosis.  

\begin{figure*}
\plotone{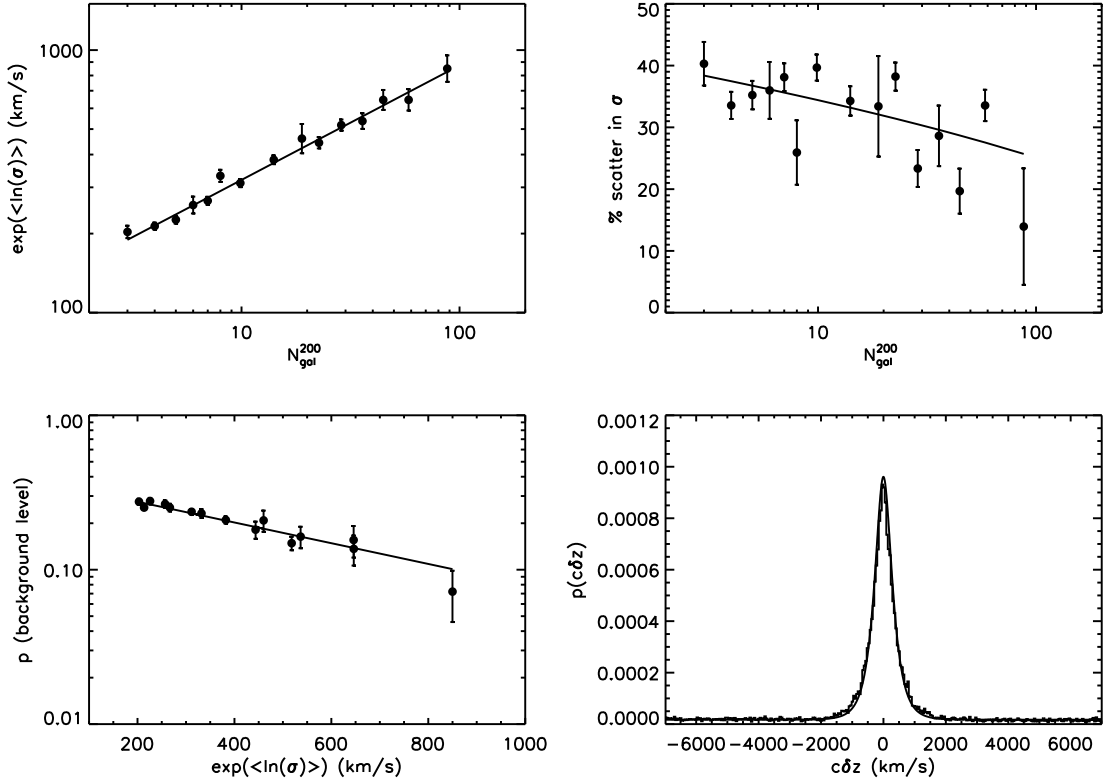}
\figcaption[f5.eps]{The mean velocity dispersion
  (i.e. $\exp{(<\!\ln{\sigma}\!>)}$), percent scatter in
  $\sigma$, and background level $p$ measured for the SDSS data.  On
  the lower right, we show the mass mixing model integrated over the
  entire data set and plotted over the cluster-weighted PVD histogram
  of the entire data set.  {\em We did not fit the stacked PVD
    histogram directly.}  The reduced chi-square between the data and
  the mass mixing model is 1.31, where we have used a robust, optimal
  bin size given by \citet{Izenman1991}.  The line fits of equations
  \ref{eqn:modmean}, \ref{eqn:scatmodeltngal}, and \ref{eqn:backmodtngal} are shown as solid lines in the lower left,
  upper left, and upper right panels respectively.  Note that we have
  plotted $100\times S$ in the upper right panel, not the linear
  relation between $S^{2}$ and $\ln{\left(N_{gals}^{200}\right)}$.\label{fig:modparms}} 
\end{figure*}

Using equations \ref{eqn:mu2disp}, \ref{eqn:normkurt},
\ref{eqn:kurttngal}, and \ref{eqn:sigmatngal}, we solve for the
parameters $<\!\ln{\sigma}\!>$ and $S^{2}$.  The background
normalization $p$ and the normalization and scatter in the
velocity-dispersion--richness relation are all modeled as power laws,
which provides a good description of the relations in both the data
and the simulations:
\begin{equation}\label{eqn:modmean}
<\!\ln{\sigma}\!>=A+B\ln{\mathrm{N_{gal}^{200}}/25}
\end{equation}
\begin{equation}\label{eqn:scatmodeltngal}
S^{2}=C+D\ln{\mathrm{N_{gal}^{200}}/25}\ .
\end{equation}
\begin{equation}\label{eqn:backmodtngal}
\ln{p}=E + F\exp{(<\!\ln{\sigma}\!>)}
\end{equation}
The fit of the measured parameters $<\!\ln{\sigma}\!>$ and $S^{2}$ to
the above relations for the maxBCG cluster
sample are shown in Figure \ref{fig:modparms}.  The parameters A, B,
C, D, E, and F are given in Table \ref{table:modparms} and the mass
mixing model values for each bin are given in Table \ref{table:modvals}.  The mean
relation plotted in this figure is corrected for BCG bias due to
improperly selected BCGs and/or BCG movement as discussed in \S
\ref{sec:bias}.  The errors for our data points are derived from the
bootstrap errors employed in the 2GAUSS method.  Note that we only have knowledge of
the full error distributions of the parameters in sets of two, and are
thus neglecting covariance between, for example, parameters A and C or
B and C, etc.  The BAYMIX method would allow for each of the three
relations to be fit simultaneously, giving full covariances between
the parameters.
 
\begin{deluxetable}{cc}
\tablewidth{0pt}
\tablecolumns{2}
\tablecaption{MaxBCG mass mixing model fit parameters.\label{table:modparms}}
\tablehead{Parameter & Value }
\startdata
mean-normalization, A        & $6.17\pm0.04$\\ 
mean-slope, B                & $0.436\pm0.015$\\ 
scatter-normalization, C     & $0.096\pm0.014$\\ 
scatter-slope, D             & $-0.0241\pm0.0050$\\ 
background-normalization, E  & $-0.980\pm0.052$\\ 
background-slope, F          & $-0.00154\pm0.00018$ 
\enddata
\end{deluxetable}

\begin{deluxetable*}{cccc}
\tablewidth{0pt}
\tablecolumns{4}
\tablecaption{MaxBCG mass mixing parameters by \ngrtwo\ bin.\label{table:modvals}}
\tablehead{Mean \ngrtwo & $<\!\ln{\sigma}\!>$ (geometric mean velocity
  dispersion) & $100\times S$ (percent scatter in $\sigma$)
  & $p$ (background level)}
\startdata
3.00 & $5.31\pm0.05$ & $40.5\pm3.5$ & $0.276\pm0.012$\\
4.00 & $5.36\pm0.03$ & $33.6\pm2.2$ & $0.252\pm0.008$\\
5.00 & $5.42\pm0.03$ & $35.3\pm2.3$ & $0.278\pm0.013$\\
6.00 & $5.55\pm0.07$ & $36.1\pm4.6$ & $0.266\pm0.015$\\
7.00 & $5.59\pm0.04$ & $38.3\pm2.2$ & $0.253\pm0.015$\\
8.00 & $5.81\pm0.05$ & $26.5\pm5.2$ & $0.232\pm0.015$\\
9.88 & $5.74\pm0.04$ & $40.0\pm2.1$ & $0.237\pm0.010$\\
14.1 & $5.95\pm0.04$ & $34.5\pm2.4$ & $0.210\pm0.011$\\
18.9 & $6.13\pm0.13$ & $33.7\pm8.1$ & $0.209\pm0.033$\\
22.7 & $6.09\pm0.05$ & $39.0\pm2.3$ & $0.187\pm0.023$\\
28.7 & $6.25\pm0.05$ & $23.5\pm3.0$ & $0.149\pm0.015$\\
35.9 & $6.29\pm0.07$ & $28.9\pm4.9$ & $0.164\pm0.026$\\
44.7 & $6.47\pm0.09$ & $20.2\pm3.6$ & $0.156\pm0.036$\\
58.4 & $6.47\pm0.09$ & $34.5\pm2.5$ & $0.137\pm0.030$\\
87.8 & $6.75\pm0.12$ & $14.9\pm9.4$ & $0.072\pm0.026$\\
\enddata
\end{deluxetable*}

\subsubsection{BCG Bias in the 2GAUSS Fitting Algorithm}\label{sec:bias}
\begin{figure}
\plotone{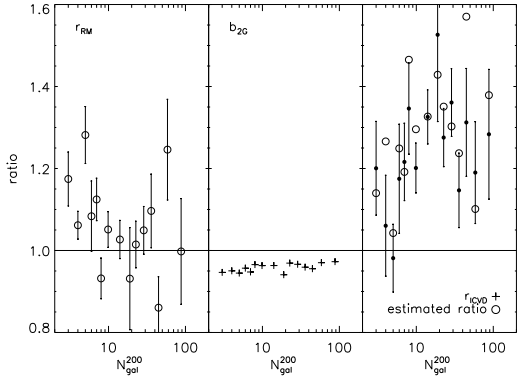}
\figcaption[f6.eps]{The bias correction to the mean velocity
  dispersion for the maxBCG clusters.  {\em Left:} The ratio of the
  random member centered dispersions to $<\!\ln{\sigma}\!>$,
  $r_{RM}$. {\em Middle:} The statistical bias in the 2GAUSS
  method, $b_{2G}$.  See Appendix \ref{app:bias} for details.  {\em
    Right:} The ratio of $<\!\ln{\sigma}\!>$ to the geometric average
  of the ICVDs for each bin in \ngrtwo, $r_{ICVD}$.  The circles in
  the right panel show the quantity
  $\sqrt{2}b_{2G}/r_{RM}$ for each bin in \ngrtwo.  Note
  that the right panel indicates
  $\sqrt{2}b_{2G}/r_{RM}\approx r_{ICVD}$.
\label{fig:sdssbias}}
\end{figure}

Despite that fact that the two mean relations plotted in Figure
\ref{fig:icvds} agree within one- to two-sigma, we show in this section
that the bias between the two relations has significance and arises
from two sources.  The first source is intrinsic statistical bias in
the 2GAUSS method itself.  Using Monte Carlo tests as described in
Appendix \ref{app:bias}, we find that this bias is approximately 3-5\%
downward and has some slight dependence on the number of data
points used in the 2GAUSS fitting method.  The Monte Carlo computation
of the bias is shown in the middle panel of Figure \ref{fig:sdssbias}.
We call this bias $b_{2G}$.  

The second source of bias is due to some combination of BCG movement with
respect to the parent halo \citep[see][]{2005MNRAS.361.1203V} and the
incorrect selection of BCGs by the maxBCG cluster detection algorithm
(i.e. mis-centering).  We can test for this effect by reconstructing
the \cgcf\ around randomly selected cluster member galaxies output
from the maxBCG cluster detection algorithm.  If the BCGs are picked
correctly and are at rest with respect to their parent halos, then by
picking a random member galaxy, we should observe the mean velocity
dispersion increase by $\sqrt{2}$.  This calculation assumes that each
stack of similar velocity dispersion clusters has a Gaussian PVD
histogram.  This test is performed in the data in the left panel of
Figure \ref{fig:sdssbias}.  We see that the random member centered
dispersions are increased above $<\!\ln{\sigma}\!>$ for each bin in
\ngrtwo, but by less than $\sqrt{2}$.  This result indicates that either
or both of the situations discussed above is happening.  The
ratio of the random member centered dispersions to $<\!\ln{\sigma}\!>$
is denoted as $r_{RM}$.

We can test the above conclusion by using the ICVDs computed in
$\S$\ref{sec:icvds}.  To do this, we calculate the ratio of
$<\!\ln{\sigma}\!>$ to the geometric average of the ICVDs for each bin
in \ngrtwo.  This ratio is plotted in the right panel of Figure
\ref{fig:sdssbias} and is called $r_{ICVD}$.  We can also estimate
this from the computations described in the previous two paragraphs.
We compute $\sqrt{2}b_{2G}/r_{RM}$ for each
bin \ngrtwo; this quantity is shown in the right panel of Figure
\ref{fig:sdssbias}.  This computation assumes that the biases add
linearly in the logarithm of the velocity dispersion.  

We find that generally $\sqrt{2}b_{2G}/r_{RM}\approx
r_{ICVD}$ within the one-sigma
errors.  This observation indicates that our explanation of the
bias observed in Figure \ref{fig:icvds} is self-consistent.  To
correct the $<\!\ln{\sigma}\!>$ values for each bin in \ngrtwo, we use
the mean of the quantity
$\sqrt{2}b_{2G}/r_{RM}$ because the ICVDs are
limited to low redshift, better sampled clusters, and our
measurements are quite noisy.

\begin{figure}
\plotone{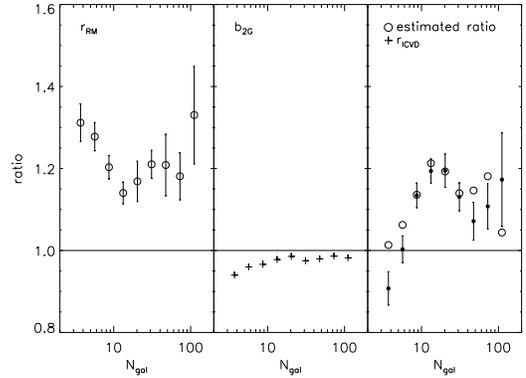}
\figcaption[f7.eps]{The bias correction to the mean velocity
  dispersion for the mock catalogs.  The panels are the same as those
  in Figure \ref{fig:sdssbias}.  According to the right panel,
  $\sqrt{2}b_{2G}/r_{RM}\approx
  r_{ICVD}$ holds in the mock catalogs as well.
\label{fig:mockbias}} 
\end{figure}

In Figure \ref{fig:mockbias}, we repeat the above computations for the
mock catalogs.  We again find that
$\sqrt{2}b_{2G}/r_{RM}\approx r_{ICVD}$ and
our explanation of the bias is self-consistent.  Furthermore, since
we know the true velocity dispersion values we can directly test our
arguments above in an absolute sense.  This comparison is discussed in
$\S$\ref{sec:mocktests}.  We find that in fact, our correction will
likely over correct the mean velocity dispersion so that it is 5-10\%
too low.  Briefly, this effect occurs because the random ``member'' we
select is in fact not always a member of the cluster.
 
Finally, the Monte Carlo tests described in Appendix \ref{app:bias} allow us to
test for bias in $S^{2}$ as well.  We find and correct for bias in
this parameter and note that on average we measure slightly lower
values of $S^{2}$ than we should, by about 5-10\%.

\subsection{Tests of the Mass Mixing Model}
We now present several checks of our method for estimating mass
mixing.  These checks fall in three broad categories.  The first set
are done with the data itself and test for self-consistency along with
dependence on sample selection functions and/or redshift.  The second
set are done with mock catalogs.  Here we run the methods developed
above on the mocks in the same way they are run on the data, and ask
whether we can recover the true velocity-dispersion--richness relation
for halos.  If the measurements on the mock catalogs do not match the
true values, then we will suspect that some of the assumptions made
above are not adequate to sufficiently describe the \cgcf\ (i.e. we
might suspect that the infall component of
the PVD histogram contributes significantly). 

The third set of tests are done with a spectroscopically-selected
catalog run on lower redshift data, the C4 catalog
\citep{2005AJ....130..968M}.  For this sample, we can compute the
distribution of velocity dispersion at fixed richness by directly
computing velocity dispersions for each individual cluster.  We can
then test our methods by comparing the measurements based on the
stacked PVD histogram to the true measured distributions.
 
\begin{figure*}
  \plotone{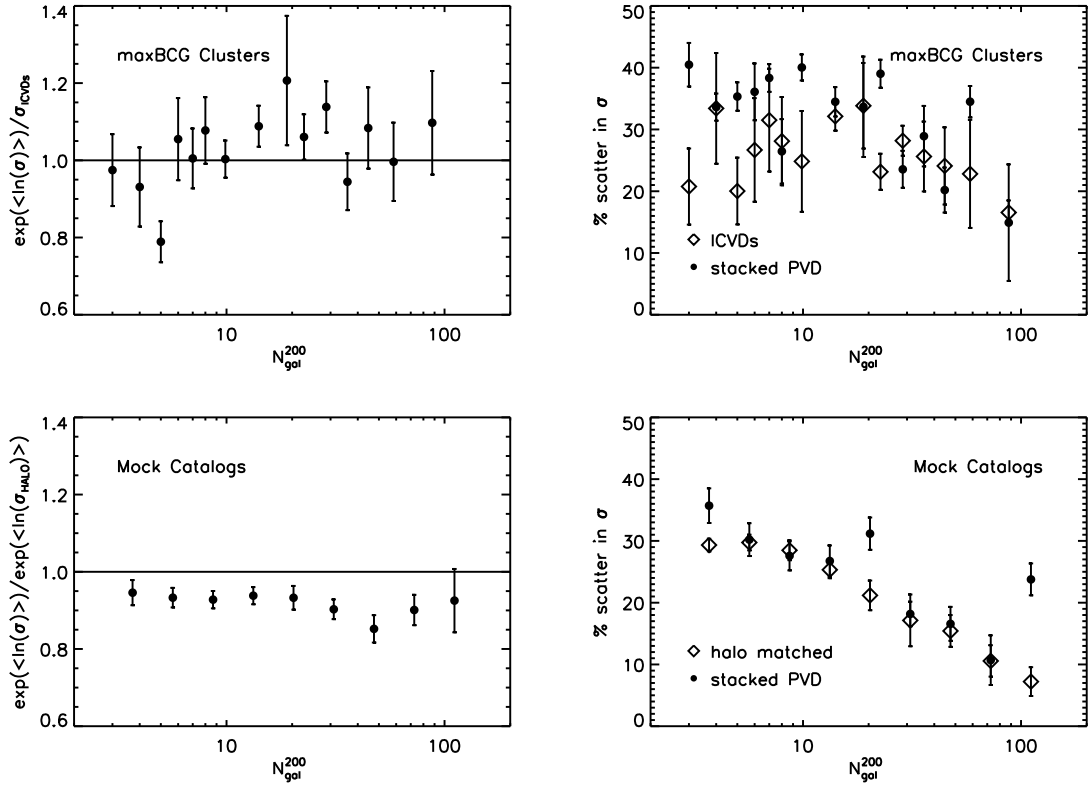} \figcaption[f8.eps]{Tests of the mass
    mixing model with the data (upper panels) and with the high
    resolution mock catalogs (lower panels).  {\em Upper Left:} The
    ratio of the geometric mean velocity dispersion determined by the
    stacked PVD histogram to that determined by the IVCDs.  {\em Upper
    Right:} The percent scatter in $\sigma$ computed directly
    from the individual cluster velocity dispersions (diamonds) to
    those computed from the stacked PVD histogram (circles).  {\em
    Lower Left:} The ratio of the geometric mean velocity dispersion
    determined by the stacked PVD histogram in the high resolution
    simulation to the true values found by matching clusters to
    halos.  {\em Lower Right:}  The percent scatter in $\sigma$
    computed using the stacked PVD histogram (circles) compared to the
    true values found by matching clusters to halos (diamonds).  The
    error bars for the simulation parameters are jackknife errors computed
    by breaking the sample into the same bins in \ngal\ as used with
    measurements of the stacked PVD histograms.  In the mock catalogs,
    note the 5-10\% downward bias of the geometric mean velocity
    dispersion, as determined by the stacked PVD histogram, with respect to
    the true values.\label{fig:mmixtest}}
\end{figure*}
 
\subsubsection{Data Dependent Tests}
As a first check of our method with the data, we look for
self-consistency.  In the lower right panel of Figure
\ref{fig:modparms}, we plot the mass mixing model integrated over the
entire data set using equations \ref{eqn:badint}, \ref{eqn:model},
\ref{eqn:modmean}, \ref{eqn:scatmodeltngal}, and
\ref{eqn:backmodtngal} on top of the full cluster-weighted stacked
PVD histogram.  \textit{We did not fit the stacked PVD histogram
  directly.}  The reduced chi-square between the data and the mass
mixing model is 1.31, where we have used a robust, optimal bin size given by
\citet{Izenman1991}.  The above model reproduces the first four moments of the
stacked PVD histogram as a function of \ngrtwo\ and reproduces the
stacked PVD histogram to a good approximation, indicating that the
model is self-consistent.

In the upper two panels of Figure \ref{fig:mmixtest}, we compare the
model parameters computed from the ICVDs (diamonds) computed using the
BISIGMA method, with those computed from the stacked PVD histogram
(circles).  The two agree to within one-sigma.  We note however that
the relation for the standard deviation of $\ln\sigma$ for the
individual cluster velocity dispersions looks ``flatter'' as function
of \ngrtwo\ than for the relation computed from the shape of the
stacked PVD histogram.

We hypothesize two possible explanations for this observation.  First, the
``flatness'' could just be a statistical fluctuation.  Notice that according
to the error bars, the relations are consistent with each other in
most instances by less than one standard deviation.  Second, the
``flatness'' could be caused by a sampling effect with the population
of clusters used to compute the individual cluster velocity
dispersions.  In other words, because we computed the individual
velocity dispersions be requiring a cluster to have ten pairs in the
\cgcf\ within three-sigma of the BCG as given by the mean velocity dispersion
relation, we selectively measure only a low redshift subset of the
cluster population.

This issue is however more than just insufficient sampling.  For small
groups of galaxies, it may be impossible to properly define an
observationally-measurable velocity dispersion unless one is willing
to stack groups of similar mass to fully sample their velocity
distributions.  Thus we hypothesize that while the two relations
disagree at low richness, the relation computed from the shape of the
stacked PVD histograms may in fact be a better indicator of scatter in
the $\sigma-$\ngrtwo\ relation for all richnesses, especially low
richness clusters.

When computing the model above, we used the entire magnitude-limited
sample of the SDSS spectroscopy.  We can investigate selection effects
by examining our model in both magnitude- and volume-limited samples.
The volume-limited samples are constructed by extracting all galaxies
above 0.4\lstar, and below the redshift at which 0.4\lstar\ is equal
to the magnitude limit of the SDSS spectroscopy.  Thus we are complete
above 0.4\lstar\, up to fiber collisions, out to this redshift.
Between the volume- and magnitude-limited samples, the differences in
the mass mixing parameters is only
slight and within the one-sigma errors.

We also binned the volume-limited sample in redshift to check for
evolution in the scatter.  Although there are only negligible
differences in the scatter in mass between between the upper and lower
redshift bins, there is a larger difference between the mean relations
for each redshift bin.  This evolution will be described in detail in
\S \ref{sec:zev} for the full magnitude-limited sample. 

Finally, we compare the mixing parameters measured with cluster
members (with redshifts) defined by the maxBCG algorithm only to those measured with
the entire spectroscopic sample (i.e. the full \cgcf).
We find no significant differences in this test.  We might suspect, as
suggested earlier, that cluster members better trace the fully
virialized regions of clusters.  Either infalling galaxies do not
contribute significantly, or the radial cut used to select members of
the \cgcf\ was small enough that most of the infalling galaxies could
be excluded, except those directly along our line-of-sight.

\subsubsection{Tests with the Mock Catalogs}\label{sec:mocktests}
After running the maxBCG cluster finder on the mock catalogs, we
measure the mass mixing of the identified clusters in the same way that
it is measured for the maxBCG clusters identified in SDSS data.  In the
bottom two panels of Figure \ref{fig:mmixtest}, the mass
mixing parameters computed using the 2GAUSS method with the mass
mixing model for clusters measured in the higher resolution
simulation are compared to the true relations, found by
matching our clusters to halos and then assigning a given cluster the
dark matter velocity dispersion of its matched halo.  We also performed the
same analysis in a lower resolution simulation.  We find that we can
successfully predict the mass mixing in both simulations above their
respective mass thresholds, except for the 5-10\% downward bias of the
mean value. 

The bias in the mean value of the velocity dispersion in the mock
catalogs is due to the imperfect selection of member galaxies by the
maxBCG cluster finding algorithm.  When we select perfectly centered
clusters (i.e. cluster in which the true BCG at rest in the halo is
found as the BCG by the maxBCG cluster finding algorithm) and repeat
the computation of $r_{RM}$, we find that the random member dispersions still
do not increase by $\sqrt{2}$.  Instead, they increase by less than
this factor and with this measured $r_{RM}$ decreasing with
\ngrtwo.  However, we can recover the factor of $\sqrt{2}$ if we use
only halo centers and only members within \rtwo\ of the halo center.
Thus, because we cannot perfectly select members, the quantity
$\sqrt{2}/r_{RM}$ is a little more than unity, so that the BCG bias
correction (in which one divides by
$\sqrt{2}b_{2G}/r_{RM}$) makes the mean too
low.  We cannot test for this effect in the data directly, but the
simulations indicate that it is less than 10\%.

The matching between clusters and halos is done according to a slight
modification of the method used by
\citet{2007astro.ph..3571R,2007astro.ph..3574R}.  The halos are first
ranked in order of richness, highest to lowest.  Then the cluster with
the most shared members with the halo is called the match.  If two
clusters share the same number of members, the one containing the halo
BCG is taken as the match.  If these two criteria fail to produce a
unique match (i.e. no cluster contains the halo's BCG), the cluster
with a the higher richness measure is chosen
as the match.  Finally, if all three criteria still fail to produce a
unique match, the matching cluster is chosen at random from all
clusters that meet all three criteria. When a match is made, both the
cluster and halo are then removed from consideration and the next
highest richness halo is matched in the same way.  This procedure
produces unique matches, but may not match every halo to a cluster or
every cluster to a halo.  Of the halos with matched clusters in the
high resolution mock catalogs, we find
that the first criteria fails in only 6.23\% of all cases.  In these
failed cases, only 5.20\%, 0.68\%, and 0.35\% of the halos are matched
using the second, third, and fourth criteria respectively.

In the SDSS data the use of cluster members only to construct the
\cgcf\ caused no change in the amount of mass mixing.  We repeat this
measurement in the higher resolution simulation using only the cluster
members selected by the maxBCG algorithm to construct the \cgcf.  We
see no significant improvement in the prediction of the true mass
mixing parameters using cluster members only as compared to using all
galaxies in the \cgcf.

In the mock catalogs, we did not properly replicate the selection
function of the SDSS spectroscopic sample.  Unfortunately, the mock
catalogs have approximately half the sky coverage area of SDSS sample, so that when
the proper selection function of the SDSS spectroscopic sample is
applied, there are too few galaxies to use with our methods.  We
require high signal-to-noise measurements of the
fourth moment of the \cgcf, which is not possible with only half the
sky coverage area.  However, we can test for the effects of spectroscopic
selection within the C4 sample, as described below.

\subsubsection{Tests with the C4 Catalog}
Using the C4 catalog \citep{2005AJ....130..968M}, we can perform an independent test
of our mass mixing method.  The C4 catalog is produced by running the C4
cluster finding algorithm on low redshift SDSS spectroscopic data.
This algorithm finds clusters using their density in 4-D color space
and 3-D position space.  We make use of five pieces of information
from the C4 catalog: a richness estimate, an estimate of the velocity
dispersion of each cluster, the BCG redshift, the mean cluster
redshift, and a ``Structure Contamination Flag'' (SCF).  This flag
takes on the values of 0, 1, or 2, depending upon the degree of
interaction of a given cluster with any of its neighbors.  Isolated
clusters have SCF$=0$ and clusters that have neighbors very close by
(i.e. $\Delta z\approx0.01$) have SCF$=2$. Clusters with SCF =$1$ are
in between these two extremes.  The mean cluster redshift is the
biweight mean \citep{1990AJ....100...32B} redshift of all SDSS
spectroscopically-sampled galaxies within 1 $h^{-1}$Mpc and $\pm0.02$
in redshift of the centroid found in the PVD histogram of the cluster.

We use every cluster in the C4 catalog with SCF$\neq 2$ and in the
redshift range $0.03<z<0.12$.  Centering on
the BCGs listed in the C4 catalog, we process the clusters in the same
way we have processed the maxBCG clusters, i.e. we measure the \cgcf\
and then apply the 2GAUSS method with the mass mixing model to the PVD
histogram.  Instead of using a projected radius cut of \rtwo\, we used
a fixed radius of 1$h^{-1}$Mpc.  We used a fixed radius here because
we have no estimate of the natural radial scaling appropriate for the
C4 richness measure.  In order to have sufficient statistics for the
computation of the mass mixing model, we are limited to splitting the
clusters into two logarithmically-spaced bins of richness.

\begin{figure}
  \plotone{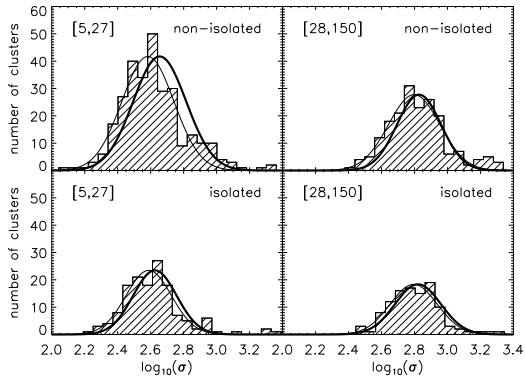} \figcaption[f9.eps]{The distribution of
    velocity dispersion in two bins of richness for the C4 catalog.
    Upper panels show the measurement for all clusters with SCF $\neq
    2$ (those with close neighbors).  Lower panels use all clusters
    with SCF $=0$ (isolated clusters).  The bold lines show the
    lognormal distributions measured by applying the 2GAUSS method
    with the mass mixing model to the C4 clusters using BCG redshifts
    in the same way as it is applied to the maxBCG clusters.  The regular
    lines are lognormal fits directly to the histograms shown
    above.  Notice that in the bottom panels, the high velocity
    dispersion tail due to C4 clusters with close neighbors is gone.\label{fig:c4mmix}}
\end{figure}

We then compare our inferred distribution of velocity dispersions with
the distribution of velocity dispersions for each individual C4
cluster in the catalog for each bin.  The results are shown in the
upper two panels of Figure \ref{fig:c4mmix}, which compares the
lognormal with our derived parameters to the best fit lognormal for
the individual C4 velocity dispersions.  Note the slight bias in the
mean between the lognormal curves computed from the mass mixing model
and the curves fit to the C4 velocity dispersions.  

\begin{figure}
  \plotone{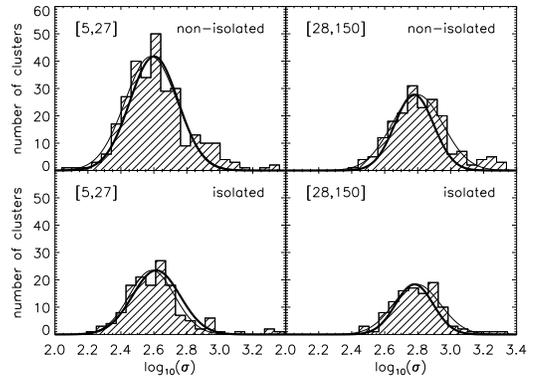} \figcaption[f10.eps]{Same as Figure
  \ref{fig:c4mmix} except the bold lines were made with the mass
  mixing model using the mean cluster redshift, not the BCG
  redshift. The geometric mean velocity dispersion as measured by the
  mass mixing model now agrees with the true geometric mean of the C4
  clusters because we have used cluster redshifts and not BCG
  redshifts.  Notice again that in the bottom panels, the high velocity
  dispersion tail due to C4 clusters with close neighbors is gone.\label{fig:c4mmixclust}}
\end{figure}

In Figure \ref{fig:c4mmixclust}, we repeat our measurements, but using
the cluster redshift instead of the BCG redshift.  In this case we see
better agreement between the mass mixing measurements and the C4
velocity dispersions.  This result indicates that the slight bias in the mean
was due to movement of the BCGs.  Finally, to be complete, we compute
the average velocity dispersion of the BCGs in each bin of C4
richness, and then use this value to correct the measurements made
using BCG centers. We find that we can reproduce the cluster redshift
measurements through this procedure. Our understanding of how
mis-centering and/or BCG movement effects our measurements is
self-consistent in both the data and mock catalogs. If we had true
cluster redshifts for the maxBCG clusters (i.e. an accurate average
redshift of all of the cluster members), then according to the results
of the C4 catalog, no correction due to BCG bias would have to be
applied to the mean velocity dispersion. 
 
We note that there seems to be high dispersion tail/shoulder in the
histograms plotted in the upper panels of Figures \ref{fig:c4mmix} and
Figures \ref{fig:c4mmixclust}.
Including clusters with SCF$=2$ increases this shoulder.  This result
is consistent with the finding of \citet{2005AJ....130..968M} that clusters with
SCF$=2$ have their measured velocity dispersions artificially
increased by their nearby neighbors.  This result is also consistent
with a finding of \citet{2007astro.ph..2241E}, that the velocity dispersions of
interacting dark matter halos form a high tail in the lognormal
distribution of velocity dispersion at fixed mass.  In the bottom
panels of Figures \ref{fig:c4mmix} and \ref{fig:c4mmixclust}, we
repeat our measurements using cluster redshifts, now including only
those clusters with SCF$=0$, (i.e. we exclude clusters with SCF$=1$ or
2).  The distribution of dispersions from this set of clusters is in
better agreement with the mass mixing method. 

In the maxBCG catalog, interacting clusters may not be as significant
of a problem because the clusters are by definition much farther away
from each other (i.e. $\Delta z \geq 0.02$ as opposed to $\Delta
z=0.01$ for some clusters in the C4 catalog).  In fact, many of the
clusters that the C4 algorithm would flag as SCF$=2$, the maxBCG
algorithm may group together.  We do not mean to imply that the maxBCG
algorithm has a significant problem of over-merging distinct objects,
but only that the redshift resolution of the cluster finder is less
than the C4 algorithm.  Thus the high velocity dispersion shoulder
would likely be down-weighted by algorithmic merging of objects
together in combination with sparse spectroscopic sampling.  For
example, if two C4 clusters with SCF$=2$ would be merged by the maxBCG
algorithm, then the velocity structure according to the maxBCG
algorithm would only be measured about a combination center, and not
two centers as in the C4 catalog.  Therefore, in combination with the
sparse spectroscopic sampling, the relative weight of these two
objects in a maxBCG PVD histogram might be decreased as compared to a
C4 PVD histogram, where they would contribute twice as much and
possibly at higher velocity dispersion.  While these arguments remain
untested at the moment, the high velocity dispersion shoulder seen in
the upper panels of Figures \ref{fig:c4mmix} and \ref{fig:c4mmixclust}
has a clear origin and is well predicted theoretically.

For the C4 sample, we can also investigate the effects the
spectroscopic selection.  We recomputed our measurements using three
higher r-band magnitude limits, 17.0, 16.5, and 16.0.  Because the
SDSS main sample r-band magnitude limit is 17.8, these three cuts
replicate increasing amounts of spectroscopic incompleteness.  We
found no statistically significant differences between these
measurements, indicating that spectroscopic incompleteness has a small
effect on our measurements for the C4 catalog.

\subsubsection{Sensitivity to the Scatter Model}
Here we investigate the sensitivity of our results to the choice of
using a lognormal to describe the scatter in \sigv\ at fixed \ngrtwo.
There may be other distributions that could possibly describe the
scatter just as well.  As a test case, we investigate how well a
Gaussian distribution describes the data in comparison with the
lognormal.

We first study whether the two scatter models result in different
conclusions.  In the SDSS data, we find that at high richness the
measured scatter assuming a Gaussian or a lognormal differ by less
than one standard deviation.  However at low richness, the measured
scatter in the two models differs by almost three standard deviations.
These results cannot tell us which model is better, but only whether
one model is equivalent to the other.  This seems to be the case in
the SDSS, except at low richness.  From a theoretical standpoint, we
prefer the lognormal model because it always ensures that
$\sigma\geq0$ without an arbitrary cutoff value.

In the C4 data the results are more dramatic: the measured scatter
between the two models differs by around eight to ten standard
deviations.  This fact primarily reflects the fact that in the C4
catalog, there is high velocity dispersion tail/shoulder, which a
lognormal distribution can fit much more easily than a Gaussian.  Such
a tail/shoulder may be less prevalent in the maxBCG clusters for reasons
discussed previously.

In the mock catalogs, the differences between the two models follow
the same differences as function of richness as seen between the two
models in the SDSS data: they are quite similar but become more
different at low richness.  Here, we can test which model provides a
better match to the intrinsic dispersion in the catalog.  We find that
the scatter derived from the Gaussian model differs from the true
scatter at around four standard deviations, whereas the scatter
derived from the lognormal model agrees well with the true scatter as
shown in the lower right panel of Figure \ref{fig:mmixtest}.  We have
explicitly verified that the distribution of velocity dispersion at 
fixed richness is approximately lognormal for the mock catalogs.  These
results give us some confidence that the mock catalogs describe the SDSS data
well and that a lognormal is a better approximation to the true
distribution than a Gaussian at all richness, but especially lower richnesses.

\subsection{The Velocity Dispersion Number Function}\label{sec:vfunct}
Using the mass mixing model and the abundance function of the maxBCG
clusters, we can integrate to find the velocity dispersion
function, usually defined as the number density of clusters per
$d\ln(\sigma)$.  This technique has been used to
find the velocity dispersion function of early-type galaxies in the
SDSS \citep{2003ApJ...594..225S} and to estimate the X-ray luminosity function of REFLEX
clusters \citep{2006ApJ...648..956S}.

In the interest of brevity, the result presented here is only
approximate.  We assume a $\Lambda$CDM cosmology for our volume
computation.  We restrict our analysis here to only those clusters in
the redshift range $0.1<z<0.3$ (over which the catalog is
approximately volume-limited), but we use the mixing results measured
from the extended catalog.  This procedure is justified because we
previously observed no change in the mixing results using the smaller
volume-limited sample.  We do not include any corrections for the
selection function of the catalog, but note that for the maxBCG
clusters the completeness and purity are at or above the 90\% level
and approximately richness independent above \ngrtwo$=10$
\citep{2007astro.ph..1265K,2007astro.ph..1268K}.  Finally, we ignore
any possible redshift evolution of the \ngrtwo\ measure.

\begin{figure*}
  \plottwo{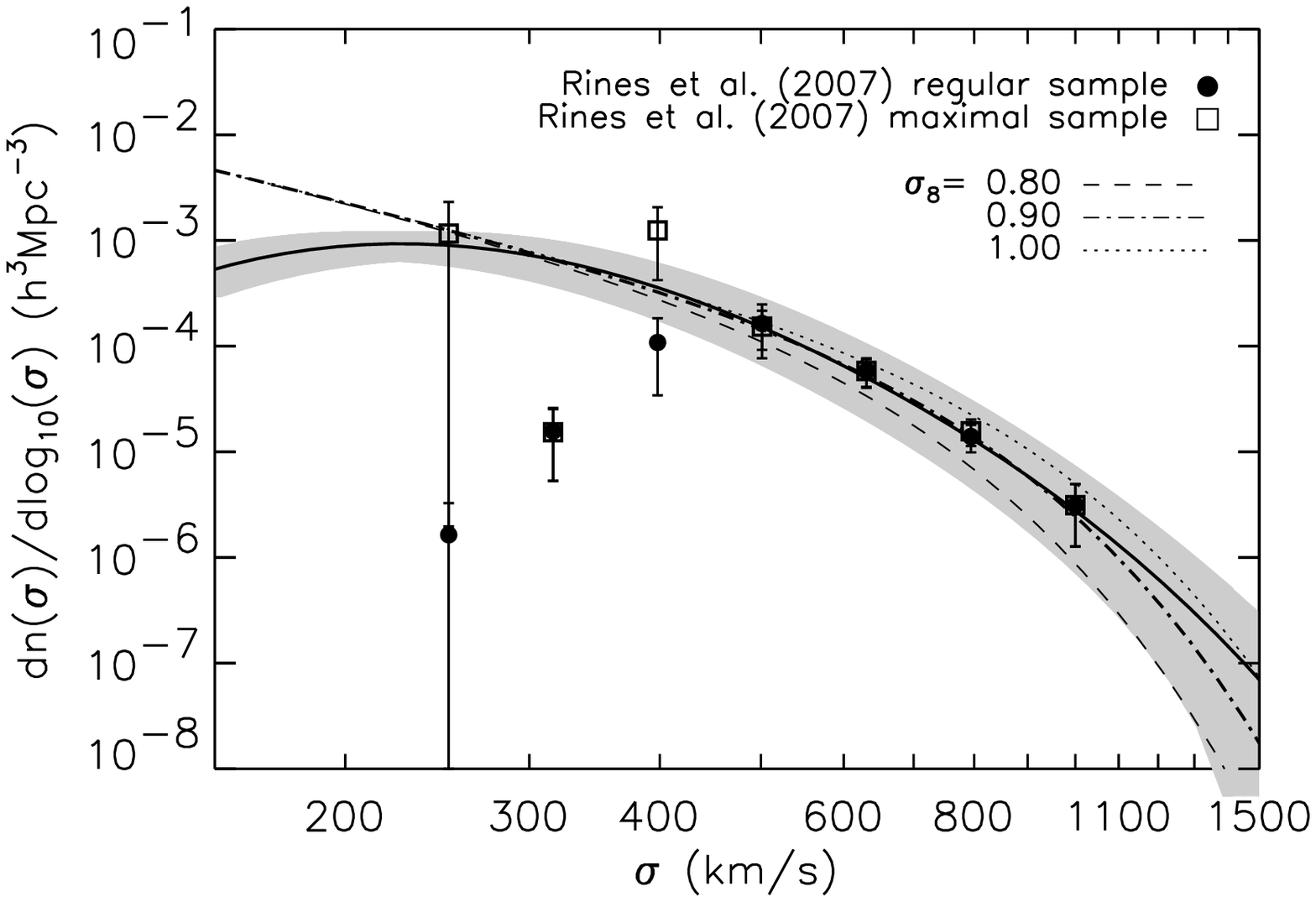}{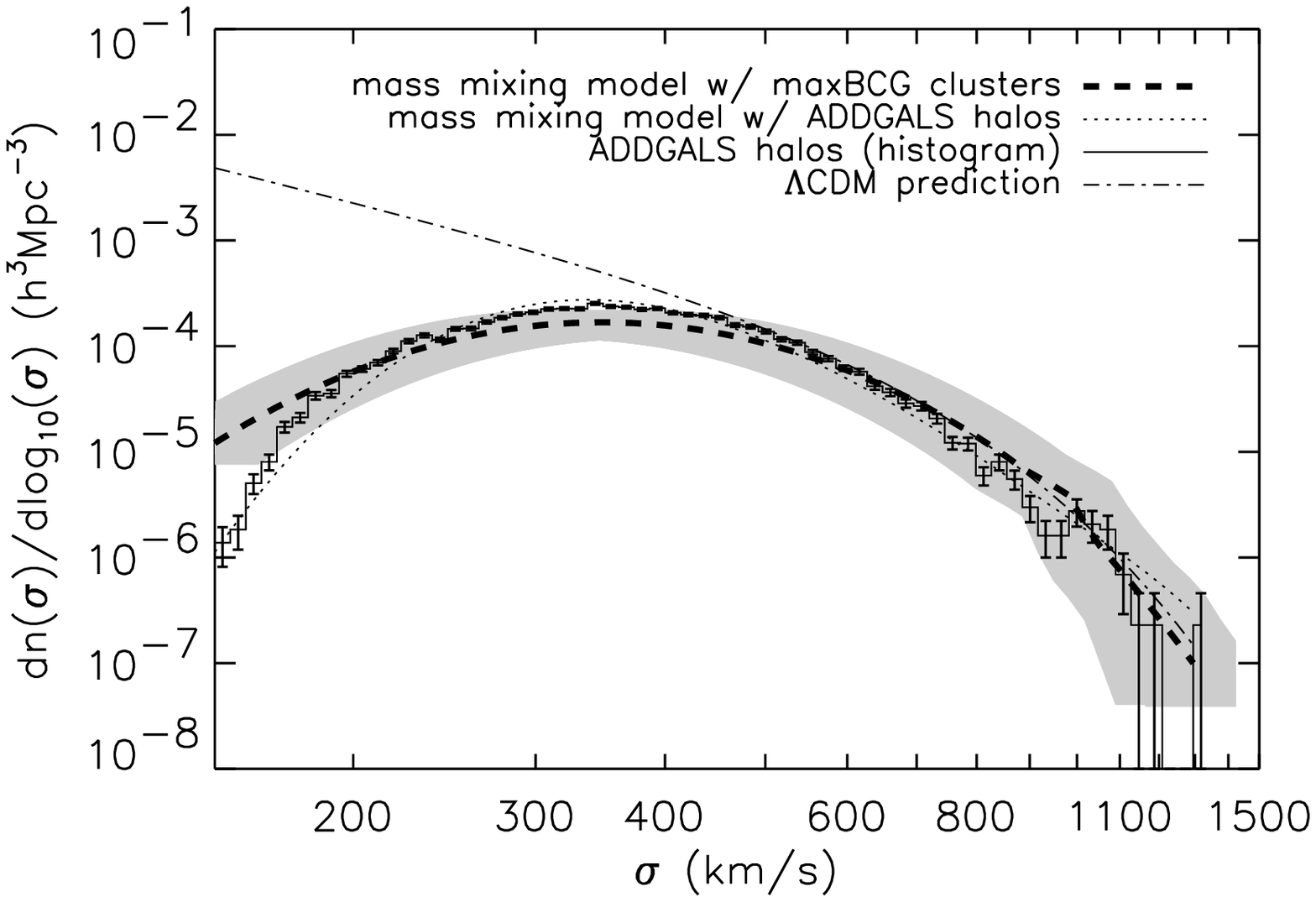}
  \figcaption[f11.eps]{The velocity dispersion number function with no
  corrections for the selection function of the maxBCG clusters (left)
  and its reconstruction in the high resolution mock catalogs (right).
  {\em Left:} The theoretical prediction for the velocity dispersion number function
  using the Jenkins mass function is combined with the N-body calibrated
  relation between mass and velocity dispersion, using the work of
  \citet{2007astro.ph..2241E}, for three values of $\sigma_8$. 
  Each assumes that $\Omega_{m}=0.30$.   The circles
  and squares show the results of \citet{2006astro.ph..6545R} for
  their regular and maximal local X-ray selected clusters samples.  The
  solid line is the velocity dispersion function of the maxBCG
  clusters.  The gray band errors indicate the systematic errors, neglecting
  any corrections for the selection function, convolved with
  the statistical errors in our measurements.  Above approximately
  1000 \kms\ the data is extrapolated.  {\em Right:} The
  dashed line with gray band errors shows the velocity dispersion
  function computed from clusters in the high resolution mock catalogs
  in exactly the same way as done with the maxBCG catalog, 
  neglecting any corrections for the selection
  function.  To estimate the systematic errors in the mass mixing
  model alone, we compute the velocity dispersion function using halo
  centers (dotted line) instead of BCGs.  The solid histogram shows the
  velocity dispersion function of all halos between redshifts of 0.1
  and 0.3 which the ADDGALS algorithm assigned three or more galaxies
  within 1 \rtwo.  This is compared with the $\Lambda$CDM
  prediction for the velocity dispersion function computed as in the
  left panel with $\sigma_{8}=0.90$ (the value in the simulation used
  for the mock catalog).\label{fig:vdispfunct}}
\end{figure*}

The velocity dispersion number function (solid curve) with systematic and
statistical errors (gray band) is given in the left panel of Figure
\ref{fig:vdispfunct}.  Due to these approximations, the systematic
errors in our result could likely be reduced in a more detailed treatment.  We aim here to
demonstrate the feasibility of such an exercise and note that much
more careful considerations of the selection function can be used to
constrain cosmology rather well \citep{2007astro.ph..3571R}.  The
systematic errors shown here arise from the selection function,
photometric redshift errors, and evolution in \ngrtwo.  We estimate
the total systematic error to be approximately 30\% in the velocity
dispersion function normalization due these effects.  We also include
a 10\% systematic error in the geometric mean velocity dispersion due
to the uncertain nature of the BCG bias correction.  The
statistical error bars are generated using a Monte Carlo technique,
assuming Poisson errors in the \ngrtwo\ number function and using the
covariance matrices of the parameters determined in the chi-square
line fits given by equations \ref{eqn:modmean} and
\ref{eqn:scatmodeltngal}.  The statistical errors are too small to be
shown alone.  Instead, we plot the systematic error convolved with the
statistical errors as a gray band in the left panel of Figure \ref{fig:vdispfunct}.

Although a detailed treatment of the velocity dispersion function,
which is beyond the scope of this paper, requires more careful
consideration of velocity bias and the systematic errors, we provide a
preliminary comparison to the predictions of the velocity dispersion
function for three values of the power spectrum normalization.  In
order to make this prediction for our sample of clusters over the
redshift range of the maxBCG catalog, $0.1<z<0.3$, we use the full
statistical relation between velocity dispersion and mass determined
in dark matter simulations by \citet{2007astro.ph..2241E}, combined
with the Jenkins mass function \citep[JMF][]{2001MNRAS.321..372J}
and its calibration for galaxy cluster surveys
\citep{2002ApJ...573....7E}.  We vary $\sigma_{8}$ between three values, 0.80, 0.90,
and 1.00, while fixing $\Omega_{m}=0.30$.  These three
curves are plotted as the dashed, dash-dotted, and dotted lines in the left
panel of Figure \ref{fig:vdispfunct}.

We can give a qualitative estimate of the effect of the selection
function of the maxBCG catalog on the velocity dispersion function.
As noted previously, because the fraction of red galaxies in a
clusters decreases with cluster mass, the maxBCG catalog may be
incomplete in the lowest mass groups.  This incompleteness would cause
our calculation to underestimate the number density of such low mass
groups, as seen in the left panel of Figure \ref{fig:vdispfunct} for
the lowest velocity dispersion groups.  Note that the low mass
deviation of the data from the predicted velocity dispersion functions occurs below
$\sigma\approx350$ \kms.  This velocity dispersion is equivalent to
\ngrtwo$\approx10$, in agreement with the determination of the
selection function by \citet{2007astro.ph..1265K,2007astro.ph..1268K}.
\citet{2007astro.ph..3571R,2007astro.ph..3574R} has shown that at high
richness, the purity of the maxBCG catalog decreases.  This decrease
in purity would cause an overestimate in the number density of the
most massive clusters, as seen in left panel of Figure
\ref{fig:vdispfunct} for the highest velocity dispersion groups.  

Also in the left panel of this figure, we compare our data to that of
\citet{2006astro.ph..6545R} who compute the velocity dispersion
function using an X-ray selected sample of local clusters.
\citet{2006astro.ph..6545R} define a regular sample which
excludes low redshift clusters and combines any multiple X-ray peaks within
the clusters into a single peak.  They also define a maximal
sample which includes all low redshift clusters and counts clusters
with multiple X-ray peaks as two objects.  We find
good agreement between our results and those of
\citet{2006astro.ph..6545R} for both samples.  Note that the
\citet{2006astro.ph..6545R} local sample of clusters has a median
redshift of 0.06 whereas our sample has median redshift closer to
0.20.  Thus we should not expect perfect agreement between the two
samples because of evolution in the mass function, but according to
the theoretical calculation, they would agree to well within 30\%.

In the right panel of Figure \ref{fig:vdispfunct}, we repeat the above
procedure using clusters in the high resolution mock catalogs with
redshifts between 0.1 and 0.3 and with $N_{gal}^{200}\geq3$.  We
make no attempts to correct for the selection function so that we can
crudely estimate its effect.  The dashed curve with the gray band
shows the results of this procedure.  In order to disentangle
systematic errors due to the selection function from those due to the
mass mixing model itself, we remake all of our measurements in the high
resolution mock catalogs using dark matter halo centers instead of the
maxBCG cluster centers; the dotted curve shows the results.  

To obtain an estimate of the true velocity dispersion function, we
additionally plot two other curves in the right panel of Figure
\ref{fig:vdispfunct}.  The solid histogram shows the true velocity dispersion
function computed from all halos with redshifts between 0.1 and 0.3
and which the ADDGALS procedure assigned three or more galaxies within
1 \rtwo.  The dash-dotted curve shows the $\Lambda$CDM
prediction for the velocity dispersion function computed as discussed
above with $\sigma_{8}=0.90$ (the value in the simulation used for the
mock catalog).  

All four curves in the right panel of Figure
\ref{fig:vdispfunct} agree to approximately within one-sigma of the
histogram error bars, above the threshold of 500 \kms.  In the mock
catalogs, this threshold corresponds to approximately $10^{14} h^{-1}M_{\odot}$ and
\ngrtwo$=10$, in agreement with the determinations of the selection
function in the mock catalogs by \citet{2007astro.ph..3571R}.  Note
also that the velocity dispersion function computed using dark matter
halo centers (dotted curve) approximately agrees with the solid histogram of ADDGALS
halos over a large range in velocity dispersion.  The systematic
errors in the mass mixing model alone are small relative to those in
the selection function.
 
\section{Dependence of the Velocity Dispersion on Secondary
  Parameters}\label{sec:tests}
In discussing the scatter model and the corrected velocity
dispersion values, we assumed no other significant
dependencies of the velocity dispersion on parameters besides \ngrtwo.
We address this assumption here through a variety of theoretically and
observationally motivated tests.
In the same way that dark matter halos are primarily characterized by
their mass, we would like to determine what parameters primarily
characterize the velocity structure of the maxBCG clusters.  By
splitting the sample of clusters at a given \ngrtwo\ value on
secondary parameters and measuring the velocity dispersion of these
secondary stacks with the 2GAUSS method, we can test for any
dependencies.

\subsection{Redshift Evolution}\label{sec:zev}
\begin{figure}[ht]
  \plotone{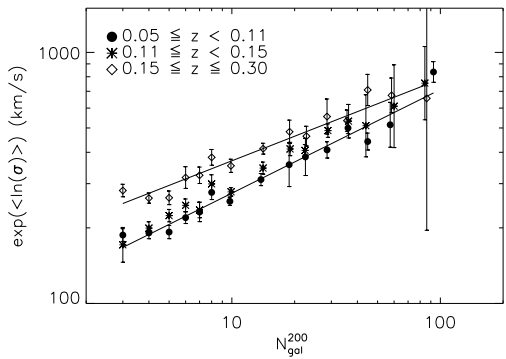} \figcaption[f12.eps]{The dependence of $\exp(<\!\ln(\sigma)\!>)$ at
    fixed richness on the BCG photometric redshift.  The dependence of the
    velocity dispersion on redshift likely indicates evolution in
    the \ngrtwo\ measure.  The solid lines are the best-fit power-laws
    to the highest and lowest redshift bins.\label{fig:vdispz}}
\end{figure}

The dependence of the velocity dispersion on redshift is shown in the
right panel of Figure \ref{fig:vdispz}.  We find a modest
dependence, with higher redshift clusters having increased velocity
dispersions over lower redshift clusters of the same richness.
Following arguments given by \citet{2007astro.ph..2241E}, we can roughly estimate if
the observed redshift dependence is due to evolution in the $\sigma-M$
relationship.
 
\citet{2007astro.ph..2241E} have found a robust relation between the velocity dispersion
and mass of dark matter halos that is constant with redshift and has
been tested with several simulation codes:
\begin{equation}
\sigma_{DM}\sim\left(h(z)M_{200c}\right)^{1/3}
\end{equation} 
where $\sigma_{DM}$ is the dark matter velocity dispersion,
$h(z)=H(z)/100$ \kms\ $Mpc^{-1}$ is the dimensionless Hubble
parameter, and $M_{200c}$ is the mass within a sphere of over density
200 times the critical density at redshift $z$.  Differentiating at
fixed mass gives
\begin{equation}
\frac{d\ln{\sigma_{DM}}}{dz}=\frac{1}{3}\frac{d\ln{h(z)}}{dz}
=\frac{h'(z)}{3h(z)}=\frac{\Omega_{m}(1+z)^{2}}{2{h(z)}^{2}}.
\end{equation}
This quantity can be computed exactly, but given the poor quality of
our data when split into three times as many bins, a rough
approximation which uses the median redshift of our sample is
sufficient.  At $z=0.2$, $h(z\!=\!0.2)\sim0.77$, which gives
$d\ln{\sigma_{DM}}/dz\approx 0.36$.  Over the redshift range of our
sample, $\Delta z\sim0.25$, so the expected change in $\ln{\sigma}$ is
$\Delta \ln{\sigma} \sim0.1$, assuming a constant velocity bias.  This
change is too small to account for the differences seen in Figure
\ref{fig:vdispz}.  Therefore we conclude that there must be evolution
in the \ngrtwo\ richness measure.  A fractional decrease in \ngrtwo\ of 30-40\% from the middle redshift bin to the upper redshift bin is consistent with our results.  Such
an evolution is likely to be a combination of true evolution in the
number of galaxies at fixed mass \citep[e.g.][]{2004ApJ...609...35K,
  2005ApJ...624..505Z} and evolution in the definition of the richness
estimator at fixed halo occupation.  There is evidence from the
evolution of richness in both the data \citep[see also][]{2007astro.ph..1265K} and the
mock catalogs, that the current definition of \ngrtwo\ does have mild
evolution.  It may however, be possible to use a slightly
modified richness estimator which does not evolve at fixed mass.  We
do not explore this possibility further here, but note that velocity
dispersion values will be useful in assessing the evolution of the
\ngrtwo\ measure and attempts to correct for it.

We finally note that the observed evolution could have other
explanations as well.  Above redshift
$\sim0.12$, the spectroscopic sample is dominated by LRGs.  A relative
velocity bias between galaxies of different colors and/or luminosities
\citep[e.g.][]{1992ApJ...396...35B} in clusters could potentially be
the cause of the observed evolution.

\subsection{Environmental Dependence and Local Density}
Considerable attention has been devoted to the environmental
dependence of the velocity dispersion in N-body simulations
\citep[e.g.][]{2001MNRAS.322..901S}.  It has been found that the
velocity dispersion does depend on local density, but only because massive halos tend
to occupy more dense environments (i.e. halo bias) and the velocity dispersion is
strongly correlated with halo mass.  No direct dependence of
the velocity dispersion on the local density has been found \citep[e.g.][]{2001MNRAS.322..901S}.

In order to test this prediction, we construct four indicators of
local density: \ngrtwo\ of closest cluster, the projected distance to
the closest cluster, the total number of cluster members of any
cluster within 5 $h^{-1}$Mpc and $\pm0.04$ in redshift, and the total number of clusters
within 5 $h^{-1}$Mpc and $\pm0.04$ in redshift.  This redshift cut is
chosen to match twice the redshift cut used in the maxBCG percolation
process to ensure that only clusters from one redshift slice on either
side of the cluster under consideration are used.

We complete the test by comparing two binning schemes.  
First, we bin in \ngrtwo\ and then on the lower 25\%, middle 50\%, and
upper 25\% quantiles of the local
density parameters within each \ngrtwo\ bin.  This method should
roughly account for the mass trend with cluster richness before
comparing local environments.  Second, we reverse the binning orders, using the
lower 25\%, middle 50\%, and upper 25\% quantiles of the \ngrtwo\
distribution in the second step.  If there
is in fact no dependence of the velocity dispersion on local density,
except because massive halos tend to occupy more dense environments,
then the relations in this binning scheme should be constant for a
given \ngrtwo\ bin.  Note however that if the halo
occupation itself correlates with local density at fixed mass, then our test could
be significantly biased.

\begin{figure*}
  \plotone{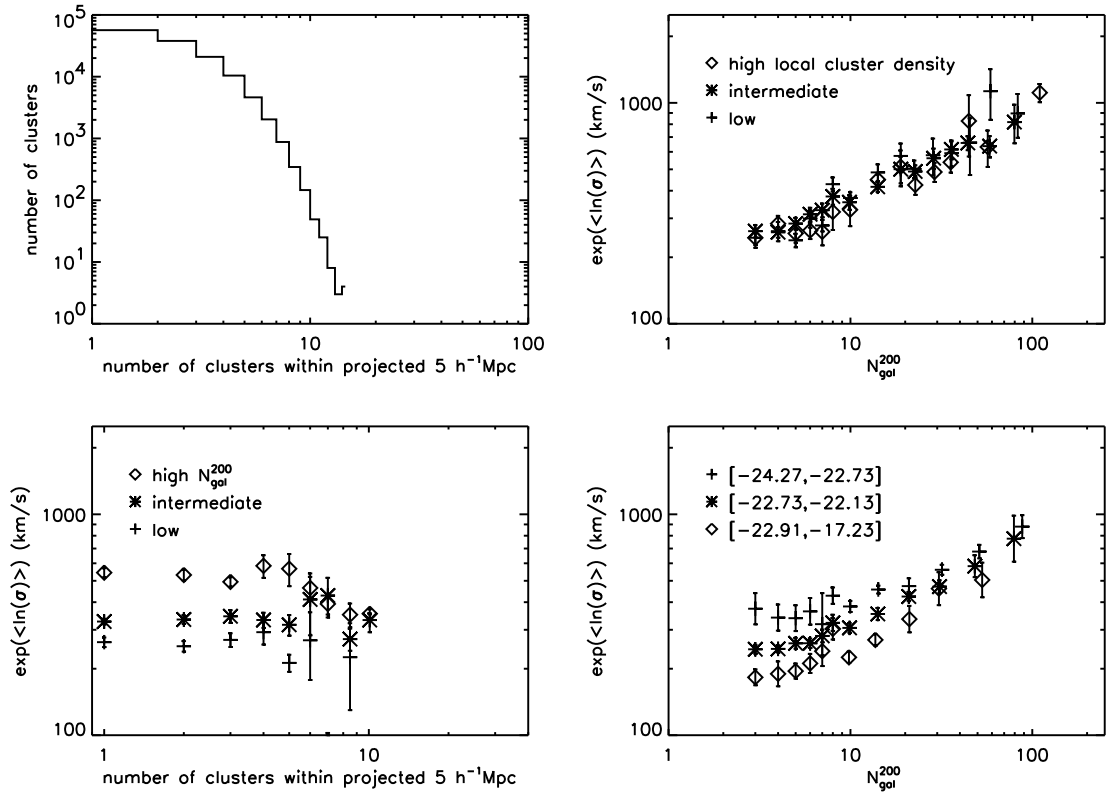} \figcaption[f13.eps]{Dependence of the
    velocity dispersion on secondary parameters.  {\em Upper left:}
    The histogram of the total number of clusters within 5 $h^{-1}$Mpc and
    $\pm0.04$ in redshift of each BCG in the \cgcf. {\em Upper right:}
    The velocity-dispersion--\ngrtwo\
     relation in bins of the number of clusters in a projected volume
    of each BCG.  {\em Lower left:} The relation between velocity dispersion
    and the number of clusters in a projected volume of BCG in bins of
    \ngrtwo.  {\em Lower right:} The dependence of the velocity
    dispersion--\ngrtwo\ relation on the absolute $i$-band magnitude
    of the BCG. No secondary parameter dependence
    would result in constant relations in the lower left panel and
    completely parallel relations with the same normalization in the
    upper right panel, assuming the halo occupation does not correlate
    with local density at fixed mass.\label{fig:secondparm}}
\end{figure*}

Using this technique, we find little significant dependence of the velocity
dispersion on any of our measures of local density.  Figure
\ref{fig:secondparm} shows the results of this 
test for one of the parameters, the total number of clusters within 5 $h^{-1}$Mpc
and $\pm0.04$ in redshift.  We have tested these results for fixed bins in
\ngrtwo\ and the local density parameters, finding that they are
robust.  We applied these tests to the mock catalogs as well, 
producing similar results.
 
\subsection{Multiple Bright Members and BCG $i$-band Luminosity}
It has been shown that clusters which have undergone recent mergers
and show significant substructure are not virialized
\citep[e.g.][]{2005PhRvE..71a6102I,1996ApJ...467...19D} and that their velocity dispersions are
increased above the expectation for their mass \citep[e.g.][]{2004A&A...425..429C,
  2004A&A...427..397H,2005A&A...442...29G}.  One might expect that a cluster with multiple bright
members that resemble the BCG has undergone a recent merger or has
significant substructure.  However, it has also been shown that dark
matter halos which form earlier at fixed mass have brighter, redder central subhalos
(i.e. brighter, redder BCGs) and lower richness
\citep[e.g.][]{2006ApJ...652...71W,2007MNRAS.374.1303C}.  Note that a
brighter BCG at fixed mass for earlier forming halos likely corresponds to a larger magnitude
difference between the BCG and the member galaxies.  

We repeat the first binning scheme used above with the $i$-band
magnitude difference of the BCG and the next brightest cluster member
as the secondary parameter.  For each bin in \ngrtwo, the brightest
member and BCG $i$-band magnitude difference distribution is split by
its lower 25\%, middle 50\%, and upper 25\% quantiles.  The naive expectation that
clusters with more than one bright member might have undergone a
recent merger or have significant substructure is not born out by the
velocity dispersions, which show no significant increase.  However,
because the computations are done at fixed richness, it could be that
late-forming halos, which have higher richness for their mass, also
have higher velocity dispersions for their mass, because they merged
recently, so that the two effects conspire to roughly cancel each other.  Unfortunately,
this hypothesis is difficult to test observationally.

One can similarly test for mass dependence of the clusters on the
luminosity of the BCG alone.  In the lower right panel of Figure
\ref{fig:secondparm}, we plot the velocity dispersion of the clusters
first binned in \ngrtwo\ and then in the absolute $i$-band magnitude
of the BCG using the lower 25\%, middle 50\%, and upper 25\% quantiles
within each \ngrtwo\ bin.  We see dependence on this parameter: at fixed richness,
clusters with more luminous BCGs have higher velocity dispersions.
This same effect is observed in the stacked X-ray measurements of
these same clusters by (Rykoff et al. 2007, in preparation); here, clusters with brighter BCGs
have on average more X-ray emission.  These observations indicate that
BCG luminosity may contain additional information about cluster mass
beyond that in \ngrtwo\ alone.  

In the case of BCG $i$-band luminosity, the expectd correlation mentioned above
consistent with the observations, since early-forming halos would have
brighter BCGs and lower richness, so that at fixed richness, halos
with brighter BCGs tend to be more massive.  However, for this
explanation to be consistent with the previous hypothesis concerning
the magnitude difference of the BCG and the next brightest cluster
member, we must assume that whether or not a halo has formed through a
major merger recently correlates more strongly with the magnitude difference of the BCG and
the next brightest member galaxy, than with the BCG $i$-band
luminosity alone.  In other words, we need to assume that the BCG
$i$-band luminosity is not indicative of a recent major merger (which
would cause the velocity dispersion to be overestimated at fixed
mass), even though BCG $i$-band luminosity correlates with formation time.

\subsection{Cluster Concentration and Radial Dependence}
Although we cannot measure the true mass concentration, we investigate the
dependence of the velocity-dispersion richness relation on the galaxy
concentration, measured here by the ratio of the number of cluster
members (determined by the maxBCG cluster finder, not the number of
pairs in the \cgcf) within 0.2 \rtwo\ to the number of members within
\rtwo.  We see no dependence of the velocity dispersions on this
parameter when the dependence on \ngrtwo\ is accounted for first.
When one bins directly on this ratio, the velocity dispersion
decreases with increasing concentration.

Finally, we investigate the dependence of the velocity dispersion
on cluster radius.  The scaling of $\sigma$ with
radius is measured in logarithmic bins of \ngrtwo.  In general, the
dispersion stays constant or decreases with radius.  This is
consistent with the results from previous studies
\citep[e.g.][]{2003AJ....126.2152R, 2006AJ....132.1275R}.

\section{Connecting Velocity Dispersion to Mass}\label{sec:massscale}
Using velocity measurements to probe cluster masses has a long history
in astronomy; the virial theorem was the earliest tool used to
determine cluster masses
\citep[e.g.][]{1933AcHPh...6..110Z,1937ApJ....86..217Z,1936ApJ....83...23S}
and remains in use today
\citep[e.g.][]{1998ApJ...505...74G,1999ApJS..125...35S,2003AJ....126.2152R,2006astro.ph..6545R,
  2006AJ....132.1275R}.
Other methods for determining cluster masses, such as the projected
mass estimator (a modified virial mass estimator)
\citep[e.g.][]{1985ApJ...298....8H,2003AJ....126.2152R,2006AJ....132.1275R},
the Jeans equation
\citep[e.g.][]{1997ApJ...478..462C,1998ApJ...505...74G,2000AJ....119.2038V,2003ApJ...585..205B,
  2003AJ....126.2152R,2004ApJ...600..657K},
and the caustic method \citep{1997ApJ...481..633D,1999MNRAS.309..610D}
have also been widely applied
\citep[e.g.][]{2003ApJ...585..205B,2003AJ....126.2152R,2006astro.ph..6545R,2005ApJ...628L..97D,
  2006AJ....132.1275R}.  

In order to connect the velocity-dispersion--richness relation to a
mass--richness relation, we use the recent results of
\citet{2007astro.ph..2241E}, who found a dark matter virial relation
which appears to hold for all redshifts and a wide range of
cosmologies.  \citet{2007astro.ph..2241E} used a suite of dissipationless
simulations run with a range of simulation codes and resolutions to
measure the velocity dispersion of dark matter particles at fixed
mass.  They find that the  
dark matter virial relation can be characterized as a power-law,
\begin{equation}
M_{200c} = 10^{15}M_{\odot}\frac{1}{h(z)}\left(\frac{\sigma_{DM}}{\sigma_{15}}\right)^{1/\alpha}
\end{equation}
where $h(z)=H(z)/100$ \kms\ $Mpc^{-1}$ is the dimensionless
Hubble parameter and $M_{200c}$ is the mass within a sphere of over
density 200 times the critical density at redshift z.  The 
values of the fit parameters for the mean relation are found to be
$\sigma_{15}=1084\pm13$ \kms\ and $\alpha=0.3359\pm0.0045$.  

\citet{2007astro.ph..2241E} additionally found that the scatter of velocity
dispersion at fixed mass is well fit by a lognormal with a small
scatter of only $0.0402\pm0.024$.  However, the lognormal scatter in
velocity dispersion at fixed mass does not directly relate to the
scatter in mass at fixed velocity dispersion without assuming the
shape of the halo mass function.  In light of this difficulty, and the
small scatter in the relation, we take the mean power-law relation
given by \citet{2007astro.ph..2241E} to be a completely deterministic relation.  

As there is still substantial theoretical uncertainty in velocity bias, this
will be a primary driver of the systematic error in the \ngrtwo-mass
relationship.  To avoid this uncertainty, we constrain a combination
of velocity bias and mass as described below.
Using the standard definition of velocity bias,
$b_{v}=\sigma_{GAL}/\sigma_{DM}$ where $\sigma_{GAL}$ is the galaxy
velocity dispersion and $\sigma_{DM}$ is the dark matter velocity
dispersions.  The virial relation for galaxy velocity dispersions 
then becomes
\begin{equation}\label{eqn:dmvt}
b_{v}^{1/\alpha}M_{200c} =
10^{15}M_{\odot}\frac{1}{h(z)}\left(\frac{\sigma_{GAL}}
{\sigma_{15}}\right)^{1/\alpha},
\end{equation}
where the quantity $b_{v}^{1/\alpha}M_{200c}$ parameterizes our lack of
knowledge about velocity bias.

To use this relation with the maxBCG clusters, we calculate
$<\!\sigma^{1/\alpha}\!>$ using the measured lognormal distribution of
$\sigma$ in each bin.  The result is
\begin{eqnarray}\label{eqn:sig3}
\lefteqn{<\!\sigma^{1/\alpha}\!>=}\nonumber\\
&&\exp{\left(
  \frac{1}{2\alpha}\ln{\mu_{(2)}^{mes}}-\frac{1}{4\alpha}\ln{\gamma_{mes}^{2}}+
  \frac{1}{8\alpha^{2}}\ln{\gamma_{mes}^{2}} \right)}
\end{eqnarray}
where $\gamma_{mes}^{2}$ and $\mu_{(2)}^{mes}$ are given by the 2GAUSS
fits for each \ngrtwo\ bin (i.e. equations \ref{eqn:kurttngal} and
\ref{eqn:sigmatngal} respectively).  This value is then substituted
for $\sigma_{GAL}$ in equation \ref{eqn:dmvt}.  To account for the
factor of $h(z)$, we repeat the 2GAUSS fits for each \ngrtwo\ bin,
weighting each pair in the \cgcf\ by $1/h(z)^{\alpha}$.  The inclusion
of this factor has a negligible effect on the observed evolution in \S
\ref{sec:zev}.  We include the correction for BCG bias by dividing our
results by the average BCG bias factor raised to the $1/\alpha$ power.

\begin{figure}
  \plotone{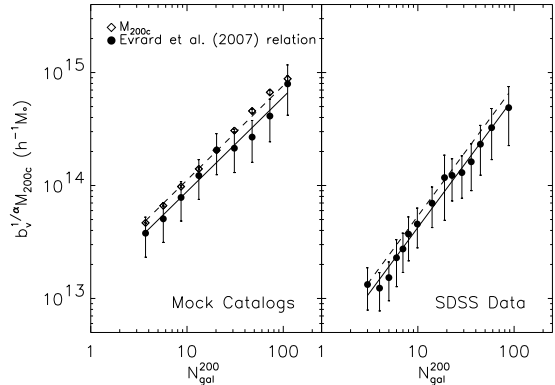} \figcaption[f14.eps]{The dark matter
    virial relation applied to our stacks of clusters in the mock
    catalogs (left, circles), and in the data on maxBCG clusters
    (right, circles).  The mean $M_{200c}$ masses for matched halos in each bin are plotted
    for the mock catalogs on the left (diamonds).  The solid line
    shows the best-fit power-law relationship between
    $b_{v}^{1/\alpha}M_{200c}$ and \ngrtwo.  The dashed lines show the
    mass normalization shifted upward by 25\% to account for the over
    correction of the BCG bias correction.
\label{fig:mass}}
\end{figure}

We apply this method first to the mock catalogs, to determine whether 
we recover an unbiased estimate of the mass--richness relation.  In the left panel
of Figure \ref{fig:mass}, we show the results of this procedure
applied to the mock catalogs.  The best-fit power-law is plotted as
the solid curve.  For comparison, the mean mass of a
cluster in each \ngrtwo\ bin, computed as the mean $M_{200c}$ mass of
the halos matched to the clusters within the bin, is shown as the
diamonds.  We again see the slight over correction of the BCG bias
correction.  Because the mass is proportional to $\sigma^{3}$, the
masses we measure in the mock catalogs are too low by $\sim 15-25\%$.
The dashed line in the left panel of Figure \ref{fig:mass} shows the
best-fit power-law relation between \ngrtwo\ and mass in the
simulation, but with the normalization increased by 25\%. Note that in
the mock catalogs the velocity bias is defined to be unity, so that
equation \ref{eqn:dmvt} should be exact with $b_{v}=1$ and
$\sigma_{GAL} = \sigma_{DM}$. 

The results for the maxBCG clusters are given in the right panel of
Figure \ref{fig:mass}.  The error bars include the theoretical
uncertainties in both $\alpha$ and $\sigma_{15}$.  The theoretical
uncertainties increase the error bars in our mass determinations by a
fixed factor uniformly across each bin.  The best-fit power-law for the
mass-\ngrtwo\ relation is
\begin{eqnarray}
\lefteqn{b_{v}^{1/\alpha}M_{200c}=} \nonumber \\
&\left(1.18^{+0.12}_{-0.11}\times10^{14}h^{-1}M_{\sun}\right)\times
\left[N_{gal}^{200}/25\right]^{1.15\pm0.12}&\ ,\nonumber
\end{eqnarray}
and is shown as the solid line in the right panel of Figure \ref{fig:mass}.
The dashed line shows the approximate effect of the BCG bias over correction by
shifting the mass normalization up by 25\%.
 
\section{Conclusions and Future Prospects}\label{sec:conclusions}
In this paper we have presented new measurements of the BCG--galaxy
velocity correlation function (BGVCF) for a sample of clusters
identified from the SDSS with the maxBCG algorithm \citep{2007astro.ph..1265K,2007astro.ph..1268K}.
Through careful modeling of the shape of the BGVCF, we have measured
the mean and scatter in velocity dispersion at fixed \ngrtwo.  We find
that the mean velocity dispersion at fixed \ngrtwo\ is well described
by a power-law.  The mean velocity dispersion increases from
$202\pm10$ \kms\ for small groups to more than $854\pm102$ \kms\ for
large clusters. The scatter in velocity dispersion at fixed \ngrtwo\
is at most $40.5\pm3.5$\% and falls to $14.9\pm9.4$\% as \ngrtwo\ increases.
We test our methods on both the C4 cluster catalog and on mock
catalogs.  Although there may be a slight 5-10\% downward bias in the
mean velocity dispersion due to the corrections made for BCG bias, our
method successfully recovers the true scatter in both of these data
sets with little bias.   

The method presented here for measuring the scatter depends on two
assumptions: (1) the Gaussianity of the PVD histogram of a stacked set of
clusters with similar velocity dispersion, and (2) the lognormal
shape of the distribution of velocity dispersion at fixed richness.
While the first assumption is valid in cluster samples produced by running the cluster
finder on realistic mock catalogs, it is hard to directly test observationally.
Simulations with galaxies based on resolved dark matter subhalos may
clarify this issue.  The second assumption is
directly supported in the mock catalogs and by independent
observations with the C4 catalog \citep{2005AJ....130..968M}.

In addition to the measurement of the mean and scatter in the
velocity-dispersion--richness relation, we explore the
dependence of the velocity dispersion on parameters secondary to
richness.  The velocity dispersion seems to be affected significantly
by the $i$-band luminosity of the BCG. We also see velocity
dispersion dependence on redshift and local density.  While the
correlation between \ngrtwo, velocity dispersion, and the BCG
$i$-band luminosity may be a true physical effect, we
interpret the correlations of \ngrtwo\ and velocity dispersion with
redshift and local density as unphysical, systematic effects of the maxBCG
cluster finder.  Ultimately, it may
be that the best way to estimate cluster mass will be to use multiple
observables in combination.  By making the comparisons of different
parameters and their dependence on velocity dispersion as done in this paper, we will
be able to determine which observables correlate significantly with mass.

Our methods, in combination with weak lensing mass profiles measured for stacked
maxBCG clusters (Sheldon et al. 2007, in preparation; Johnston et
al. 2007, in preparation) and the radial phase-space information contained in the \cgcf, will allow
for precise determinations of the velocity bias and the anisotropy of
galaxy orbits in clusters.  Precise measurements of these quantities
will help to constrain current theoretical models
of galaxy clustering and the velocity bias between dark matter and galaxies.

This work also demonstrates the feasibility of using our methods to measure the velocity
dispersion function.  The velocity dispersion
function computed in this paper agrees with the results of
\citet{2006astro.ph..6545R}.  However, given the current estimated
systematic errors in our computation (due the
selection function, photometric redshift errors, evolution in
\ngrtwo, and the BCG bias correction), we are unable to reach any strong conclusions about
the magnitude of $\sigma_{8}$.  With this caveat, we do however
see from Figure \ref{fig:vdispfunct}, that our results support a
higher value of $\sigma_{8}$ than the most recent CMB+LSS estimates
\citep[e.g.][]{2006astro.ph..3449S,2006PhRvD..74l3507T}, as recently suggested by other
analyses \citep[e.g.][]{2006astro.ph.10135B,2007astro.ph..2241E,2007astro.ph..3571R}.
This conclusion is however degenerate with velocity bias.  A velocity
bias of approximately 1.1-1.2 could equally well explain our results.

The methods presented in this paper are a significant advancement for
the use of optical cluster
surveys to determine cosmology.  Our method can fully characterize the
velocity-dispersion--richness relation for any optical cluster survey with a
large spectroscopic sample.  Future redshift surveys with more galaxy
redshifts will allow for more precise measurements of this scatter.
Specifically, because the SDSS spectroscopy is mostly at or below
$z\sim0.1$, a higher redshift sample of spectroscopy would allow for
further tests of any redshift dependence.

The measurement of the scatter in mass--observable relations is key to
the measurement of cosmology from galaxy cluster surveys and
self-calibration schemes
\citep{2004PhRvD..70d3504L,2005PhRvD..72d3006L}.  Through adding an
additional piece of observational
information, the methods developed here will undoubtedly tighten
constraints on and lift degeneracies in current estimates of
cosmology.



\acknowledgments 
MRB would like to thank Brian Nord, Eli Rykoff, and Jiangang Hao for the useful discussions and suggestions.  We also thank Mike Warren for running the higher resolution simulation described here, and Ken Rines for the generous use of his data.  MRB acknowledges the support of the Michigan Space Grant Consortium.  TAM acknowledges support from NSF grant AST 044327 and 0206277.  RHW was partially supported by NASA through Hubble Fellowship grant HST-HF-01168.01-A awarded by the Space Telescope Science Institute, and also received support from the U.S. Department of Energy under contract number DE-AC02-76SF00515.  AEE acknowledges support from the Miller Foundation for Basic Research in Science at University of California, Berkeley.  The research described in this paper was performed in part at the Jet Propulsion Laboratory, California Institute of Technology, under a contract with the National Aeronautics and Space Administration.  The authors acknowledge the partial support of the Michigan Center for Theoretical Physics.

The Sloan Digital Sky Survey (SDSS) is a joint project of The
University of Chicago, Fermilab, the Institute for Advanced Study, the
Japan Participation Group, The Johns Hopkins University, the
Max-Planck-Institute for Astronomy, Princeton University, the United
States Naval Observatory, and the University of Washington. Apache
Point Observatory, site of the SDSS, is operated by the Astrophysical
Research Consortium. Funding for the project has been provided by the
Alfred P. Sloan Foundation, the SDSS member institutions, the National
Aeronautics and Space Administration, the National Science Foundation,
the U.S. Department of Energy, and Monbusho. The SDSS Web site is
http://www.sdss.org/.  This work made extensive use of the NASA
Astrophysics Data System and of the {\tt astro-ph} preprint archive at
{\tt arXiv.org}.

\appendix

\section{Group Weighted EM-Algorithm for 1-Dimensional Gaussian
  Mixtures with Equal Means}\label{app:emalg}

Here we present the modification of the standard EM-algorithm \citep{dm77} for
Gaussian mixture models that is used to fit the PVD histograms.  It
has the advantage of weighting clusters (or groups) of points evenly
and fixing the the mean of each Gaussian to be equal.  We also verify
that the algorithm works using simple numerical experiments.  The
derivation and notation given here is that of \citet{2000astro.ph..8187C} with our
changes noted appropriately.\\

Let $j$ index the number of Gaussians in the model, $i$ index the data
points, and $N_{i}$ be the total number of data points in the group
from which the $ith$ data point is drawn.  The statistical model for
the entire set data set will be a sum of Gaussians plus a single
constant background component.  We differ from \citet{2000astro.ph..8187C} by
presenting the derivation of the algorithm with the background
components included.  Letting the Gaussian components be given as
\begin{equation}
        \label{eqn:gauss}
        \phi(x,\mu,\sigma_{j})=\frac{1}{\sigma_{j}\sqrt{2\pi}}\,\exp\left({\frac{-(x-\mu)^2}{2\sigma_{j}^{2}}}\right)
\end{equation}
where $\mu$ is the common mean of all of the model components, $\sigma_{j}$ is the standard deviation of the $jth$ Gaussian, and the background component, $U(x)$, be given as
\begin{equation}
        \label{eqn:uniform}
                U(x)=\frac{1}{2L}
\end{equation}
where $2L$ is the range of all of the data, $X$, we can write the model as
\begin{equation}
\label{eqn:clustTmod}
P\left(X\,|\,\mu,\sigma_{1},\ldots,\sigma_{j}\right)=p_{0}\,U(x)+\sum_{j=1}{p_{j}\,\phi(x,\mu,\sigma_{j})}
\end{equation}
where the $p_{j}$'s are the weights for each model component and we
require that $p_{0}+\sum_{j}{p_{j}}=1$.  This model is exactly that of
the standard EM-algorithm for Gaussian mixtures.  The difference here will be in the structure of the latent variables.\\

Let $z_{ij}$ be defined such that $z_{ij}=1$ if the $ith$ data point is in the $jth$ Gaussian and $z_{ij}=0$ otherwise.  Now we can write the complete data log-likelihood as
\begin{equation}
\label{eq:cdll}
{\cal L}=\sum_{i}{z_{i0}\ln{\left(p_{0}\,U(x)\right)}}+\sum_{j}{\sum_{i}{z_{ij}\left[\ln{p_{j}}+\ln{\phi(x,\mu,\sigma_{j})}\right]}}.
\end{equation}

Now we perform the expectation step of the algorithm given the current parameter guesses, $\tilde{\theta}$, computing
\begin{eqnarray}
\label{eq:Qdef}
Q&=&E({\cal L}\,|\,X,\tilde{\theta})\nonumber\\
 &=&\sum_{i}{E(z_{i0}\,|\,X,\tilde{\theta})\ln{\left(p_{0}\,U(x)\right)}}+\sum_{j}{\sum_{i}{E(z_{ij}\,|\,X,\tilde{\theta})\left[\ln{p_{j}}+\ln{\phi(x,\mu,\sigma_{j})}\right]}}\nonumber\\
 &=&\sum_{i}{\tau_{i0}\ln{\left(p_{0}\,U(x)\right)}}+\sum_{j}{\sum_{i}{\tau_{ij}\left[\ln{p_{j}}+\ln{\phi(x,\mu,\sigma_{j})}\right]}}\nonumber
\end{eqnarray}
where $\tau_{ij}$ is now weighted by cluster (or group) to give
\begin{eqnarray}
\label{eq:tauij}
\tau_{ij}&=&E(z_{ij}\,|\,X,\tilde{\theta}\,)\nonumber\\
 &=&P(x_{i}{\rm \ is\ in\ group \ j}\,|\,X ,\tilde{\theta})\nonumber\\
 &=&\frac{1}{N_{i}}\frac{P(x_i\,|\, {\rm group\ j })P( {\rm group\ }j)}{\sum_r P(x_i\,|\, {\rm group\ r })P( {\rm group\ }r)}\nonumber\\
 &=&\frac{1}{N_{i}}\frac{\tilde{p}_{j}\,\phi(x_i,\tilde{\mu},\tilde{\sigma}_{j})}{\tilde{p}_{0}\,U(x)+\sum_r \tilde{p}_{r}\,\phi(x_i,\tilde{\mu},\tilde{\sigma}_{r})}
\end{eqnarray}
and for the background
\begin{eqnarray}
\label{eq:taui0}
\tau_{i0}&=&E(z_{i0}\,|\,X,\tilde{\theta}\,)\nonumber\\
 &=&P(x_{i}{\rm \ is\ in\ background}\,|\,X ,\tilde{\theta})\nonumber\\
 &=&\frac{1}{N_{i}}\frac{P(x_i\,|\, {\rm background })P( {\rm background})}{\sum_r P(x_i\,|\, {\rm group\ r })P( {\rm group\ }r)}\nonumber\\
 &=&\frac{1}{N_{i}}\frac{\tilde{p}_{0}\,U(x)}{\tilde{p}_{0}\,U(x)+\sum_{r}{\tilde{p}_{r}\,\phi(x_i,\tilde{\mu},\tilde{\sigma}_{r})}}.
\end{eqnarray}\\

Now we compute the maximization step by maximizing $Q$ over the
parameters.  Following \citet{2000astro.ph..8187C}, we first rewrite
$Q$ into a more manageable form and then maximize $Q$.  Using the definition of our model and its components we have
\begin{equation}
Q=\sum_{i}{\tau_{i0}\,[\ln{p_{0}}-\ln{(2L)}]}+\sum_{j}{\sum_{i}{\tau_{ij}\left[\ln{p_{j}}-\frac{1}{2}\ln{2\pi}-\ln{\sigma_{j}}-\frac{(x_{i}-\mu)^{2}}{2\sigma_{j}^{2}}\right]}}.
\end{equation}
Letting
\begin{equation}
\label{eq:muhat}
\hat{\mu}=\frac{\sum_{ij}{{\tau_{ij}\,x_{i}}}}{\sum_{ij}{{\tau_{ij}}}}
\end{equation}
and noting that $\displaystyle\sum_{ij}{\frac{\tau_{ij}}{\sigma_{j}^{2}}(x_{i}-\hat{\mu})(\hat{\mu}-\mu)}=0$, we get
\begin{eqnarray}
\sum_{ij}{\tau_{ij}(x_{i}-\mu)^{2}}&=&\sum_{ij}{\tau_{ij}\left[(x_{i}-\hat{\mu})+(\hat{\mu}-\mu)\right]^{2}}\nonumber\\
&=&\sum_{ij}{\tau_{ij}\left[(x_{i}-\hat{\mu})^{2}+(\hat{\mu}-\mu)^{2}\right]}\nonumber\\
&=&\sum_{j}{{\cal B}_{j}}\,+\,\sum_{ij}{\tau_{ij}(\hat{\mu}-\mu)^{2}}\nonumber,
\end{eqnarray}
where
\begin{equation}
{\cal B}_{j}=\sum_{i}{\tau_{ij}(x_{i}-\hat{\mu})^{2}}.
\end{equation}
Now we rewrite $Q$ as
\begin{equation}
\label{eq:fullQ}
Q=\sum_{i}{\tau_{i0}\,[\ln{p_{0}}-\ln{(2L)}]}+\sum_{j}{\sum_{i}{\tau_{ij}\left[\ln{p_{j}}-\frac{1}{2}\ln{2\pi}-\ln{\sigma_{j}}-\frac{{\cal B}_{j}}{2\sigma_{j}^{2}}-\frac{(\hat{\mu}-\mu)^{2}}{1\sigma_{j}^{2}}\right]}}.
\end{equation}
The maximum of $Q$ subject to $p_{0}+\sum_{j}{p_{j}}=1$ will be given by letting $\mu$ be given by equation \ref{eq:muhat} and the rest of the parameters by
\begin{equation}
\label{eq:phat}
\hat{p_{j}}=\frac{\sum_{i}{\tau_{ij}}}{{\sum_{ij}{\tau_{ij}}}}, 
\hat{\sigma}_{j}^{2}=\frac{{\cal B}_{j}}{\sum_{ij}{\tau_{ij}}}.
\end{equation}\\

The primary differences between the derivation given here and that
given by \citet{2000astro.ph..8187C} are the re-weighted expectation values of
$z_{ij}$, equations \ref{eq:tauij} \& \ref{eq:taui0}, and that the
means of each of the Gaussian components are fixed to one value
(i.e. equation \ref{eq:muhat}).  The cluster-weighting is encoded in
the factor of $1/N_{i}$ in equations \ref{eq:tauij} and
\ref{eq:taui0}.  Each cluster of points contributes an equal amount to
the parameters of the mixture model, making the total model wighted by
cluster, not by point.   

\begin{figure}
  \plotone{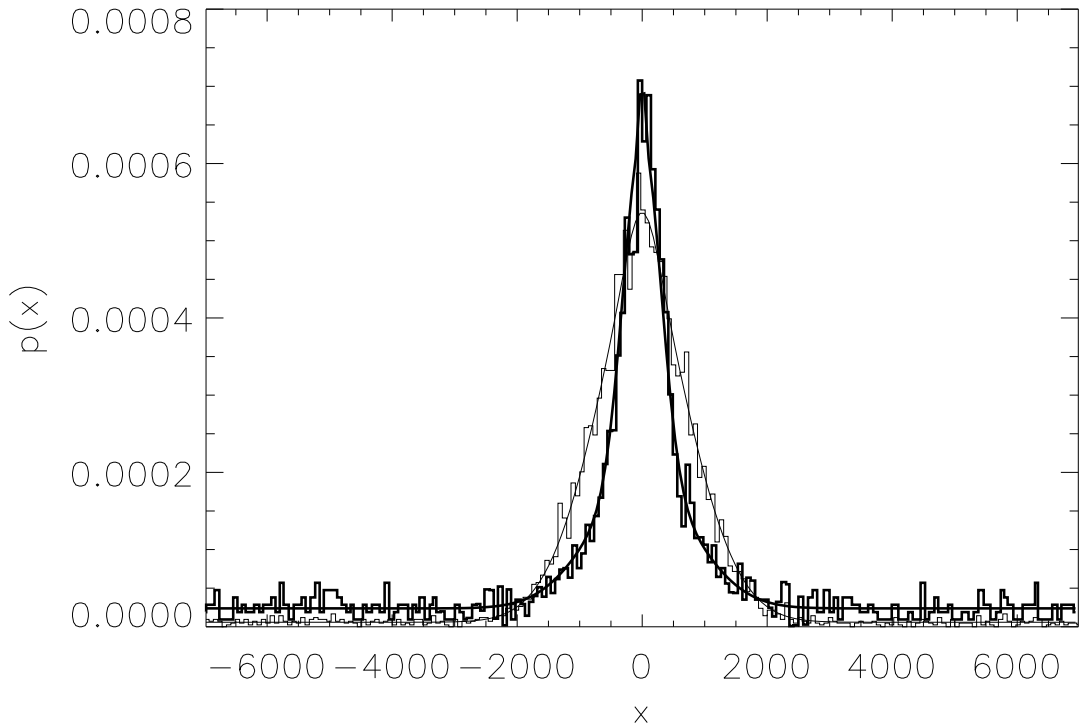}
  \figcaption[f15.eps]{A simple test of the cluster-weighted
    EM algorithm.  The bold line and histogram are cluster-weighted;
    the thin line and histogram are point-weighted.  The
    point-weighted histogram can statistically ``hide''
    non-Gaussianity because it allows one component to contribute the majority
    of the points to the histogram, and thus determine its shape.  The
    cluster-weighted histogram eliminates this issue by forcing each
    component to contribute equally.  See text for
    details. 
\label{fig:emclusttestfig}}
\end{figure}

In Figure \ref{fig:emclusttestfig}, we plot the point-weighted
histogram of data created from 5000 and 500 random draws from two
Gaussians with dispersion 800 and 300 respectively along with the
standard EM algorithm fit using 10 components.  We also plot the
cluster-weighted histogram of the same data fit with the EM algorithm
derived above using 10 components (bold line).  The shapes seen in the
figure are expected.  In the point-weighted case, the 5000 samples of
the 800 width Gaussian dominate the PVD histogram so it appears to be
Gaussian.  However, in the cluster-weighted case, the two Gaussians of
width 800 and 300 contribute equally.  Thus we explicitly see the the
non-Gaussianity in the histogram.  In effect, by not weighting the
histogram by cluster, one can ``hide'' non-Gaussianity statistically.
 
\section{Statistical Bias in the 2GAUSS Method}\label{app:bias}
In this appendix, we describe the statistical bias correction we apply
to the 2GAUSS method.  Suppose we have random samples,
$\mathbf{D}=\{d_{1},d_{2},\ldots,d_{N}\}$, from a distribution,
$p(x)$, characterized by a set of parameters, $\mathbf{
  z}=\{z_{1},z_{2},\ldots,z_{n}\}$, so that we can write
$p(x;\mathbf{z})$.  A simple example would be a Gaussian characterized
by its mean and variance.  From these random samples, we can construct
estimators, $\hat{\mu}$, of the moments of the distribution,
$\mu_{(i)} = \int_{-\infty}^{\infty}{x^{i}p(x;\mathbf{z})dx}$.  A well
known example of an estimator of $\mu_{(1)}$ is the mean,
$\hat{\mu}=\sum_{k=0}^{N}{d_{k}}/N$.  An estimator of the $ith$ moment
of a distribution is said to be unbiased if
$\int_{-\infty}^{\infty}{\hat{\mu}(\mathbf{D})\,p(x;\mathbf{z})dx}=\mu_{(i)}$.
In some cases the sampling distribution of an estimator can be
computed exactly, so that bias can be computed and corrected for analytically.

Because the 2GAUSS method is rather complicated, instead of attempting
to compute the bias analytically, we use a Monte Carlo method to
estimate the bias.  For each bin in \ngrtwo, we construct 10,000 Monte
Carlo samples of the set of velocity separation values, $\mathbf{v}=\{v_{1},v_
{2},\ldots,v_{N}\}$, using the measured parameters $<\!\ln{\sigma}\!>$
and $S^{2}$, the number of clusters in the bin, and the number of
samples per cluster in the bin.  Then we remeasure $<\!\ln{\sigma}\!>$
and $S^{2}$ for each Monte Carlo sample.  From these 10,000
reestimations of $<\!\ln{\sigma}\!>$ and $S^{2}$, we estimate the bias
in the 2GAUSS method.  We then use this bias to correct our measurements. 

We note here that the Monte Carlo tests indicate that the estimations
of $<\!\ln{\sigma}\!>$ and $S^{2}$ are correlated in a non-trivial
way.  The BAYMIX method would naturally account for these correlations
in a transparent way.  Alternatively, a different, unbiased procedure
could be used to estimate $<\!\ln{\sigma}\!>$ and $S^{2}$.  One
candidate method might be the use of Gauss-Hermite moments
\citep{1993ApJ...407..525V}.  Given that we already have a well
developed method to estimate $<\!\ln{\sigma}\!>$ and $S^{2}$, we do
not explore this possibility here.  A detailed understanding of the
correlations is beyond the scope of this work, but would be necessary
for the use of these results to understand cosmology precisely.





\begin{thebibliography}{97}
\expandafter\ifx\csname natexlab\endcsname\relax\def\natexlab#1{#1}\fi

\bibitem[{{Abazajian} {et~al.}(2004){Abazajian}, {Adelman-McCarthy},
  {Ag{\"u}eros}, {Allam}, {Anderson}, {Anderson}, {Annis}, {Bahcall}, {Baldry},
  {Bastian}, {Berlind}, {Bernardi}, {Blanton}, {Bochanski}, {Boroski},
  {Briggs}, {Brinkmann}, {Brunner}, {Budav{\'a}ri}, {Carey}, {Carliles},
  {Castander}, {Connolly}, {Csabai}, {Doi}, {Dong}, {Eisenstein}, {Evans},
  {Fan}, {Finkbeiner}, {Friedman}, {Frieman}, {Fukugita}, {Gal}, {Gillespie},
  {Glazebrook}, {Gray}, {Grebel}, {Gunn}, {Gurbani}, {Hall}, {Hamabe},
  {Harris}, {Harris}, {Harvanek}, {Heckman}, {Hendry}, {Hennessy}, {Hindsley},
  {Hogan}, {Hogg}, {Holmgren}, {Ichikawa}, {Ichikawa}, {Ivezi{\'c}}, {Jester},
  {Johnston}, {Jorgensen}, {Kent}, {Kleinman}, {Knapp}, {Kniazev}, {Kron},
  {Krzesinski}, {Kunszt}, {Kuropatkin}, {Lamb}, {Lampeitl}, {Lee}, {Leger},
  {Li}, {Lin}, {Loh}, {Long}, {Loveday}, {Lupton}, {Malik}, {Margon},
  {Matsubara}, {McGehee}, {McKay}, {Meiksin}, {Munn}, {Nakajima}, {Nash},
  {Neilsen}, {Newberg}, {Newman}, {Nichol}, {Nicinski}, {Nieto-Santisteban},
  {Nitta}, {Okamura}, {O'Mullane}, {Ostriker}, {Owen}, {Padmanabhan},
  {Peoples}, {Pier}, {Pope}, {Quinn}, {Richards}, {Richmond}, {Rix}, {Rockosi},
  {Schlegel}, {Schneider}, {Scranton}, {Sekiguchi}, {Seljak}, {Sergey},
  {Sesar}, {Sheldon}, {Shimasaku}, {Siegmund}, {Silvestri}, {Smith}, {Smol{\v
  c}i{\'c}}, {Snedden}, {Stebbins}, {Stoughton}, {Strauss}, {SubbaRao},
  {Szalay}, {Szapudi}, {Szkody}, {Szokoly}, {Tegmark}, {Teodoro}, {Thakar},
  {Tremonti}, {Tucker}, {Uomoto}, {Vanden Berk}, {Vandenberg}, {Vogeley},
  {Voges}, {Vogt}, {Walkowicz}, {Wang}, {Weinberg}, {West}, {White}, {Wilhite},
  {Xu}, {Yanny}, {Yasuda}, {Yip}, {Yocum}, {York}, {Zehavi}, {Zibetti}, \&
  {Zucker}}]{2004AJ....128..502A}
{Abazajian}, K. {et~al.} 2004, \aj, 128, 502

\bibitem[{{Abazajian} {et~al.}(2005){Abazajian}, {Adelman-McCarthy},
  {Ag{\"u}eros}, {Allam}, {Anderson}, {Anderson}, {Annis}, {Bahcall}, {Baldry},
  {Bastian}, {Berlind}, {Bernardi}, {Blanton}, {Bochanski}, {Boroski},
  {Brewington}, {Briggs}, {Brinkmann}, {Brunner}, {Budav{\'a}ri}, {Carey},
  {Castander}, {Connolly}, {Covey}, {Csabai}, {Dalcanton}, {Doi}, {Dong},
  {Eisenstein}, {Evans}, {Fan}, {Finkbeiner}, {Friedman}, {Frieman},
  {Fukugita}, {Gillespie}, {Glazebrook}, {Gray}, {Grebel}, {Gunn}, {Gurbani},
  {Hall}, {Hamabe}, {Harbeck}, {Harris}, {Harris}, {Harvanek}, {Hawley},
  {Hayes}, {Heckman}, {Hendry}, {Hennessy}, {Hindsley}, {Hogan}, {Hogg},
  {Holmgren}, {Holtzman}, {Ichikawa}, {Ichikawa}, {Ivezi{\'c}}, {Jester},
  {Johnston}, {Jorgensen}, {Juri{\'c}}, {Kent}, {Kleinman}, {Knapp}, {Kniazev},
  {Kron}, {Krzesinski}, {Lamb}, {Lampeitl}, {Lee}, {Lin}, {Long}, {Loveday},
  {Lupton}, {Mannery}, {Margon}, {Mart{\'{\i}}nez-Delgado}, {Matsubara},
  {McGehee}, {McKay}, {Meiksin}, {M{\'e}nard}, {Munn}, {Nash}, {Neilsen},
  {Newberg}, {Newman}, {Nichol}, {Nicinski}, {Nieto-Santisteban}, {Nitta},
  {Okamura}, {O'Mullane}, {Owen}, {Padmanabhan}, {Pauls}, {Peoples}, {Pier},
  {Pope}, {Pourbaix}, {Quinn}, {Raddick}, {Richards}, {Richmond}, {Rix},
  {Rockosi}, {Schlegel}, {Schneider}, {Schroeder}, {Scranton}, {Sekiguchi},
  {Sheldon}, {Shimasaku}, {Silvestri}, {Smith}, {Smol{\v c}i{\'c}}, {Snedden},
  {Stebbins}, {Stoughton}, {Strauss}, {SubbaRao}, {Szalay}, {Szapudi},
  {Szkody}, {Szokoly}, {Tegmark}, {Teodoro}, {Thakar}, {Tremonti}, {Tucker},
  {Uomoto}, {Vanden Berk}, {Vandenberg}, {Vogeley}, {Voges}, {Vogt},
  {Walkowicz}, {Wang}, {Weinberg}, {West}, {White}, {Wilhite}, {Xu}, {Yanny},
  {Yasuda}, {Yip}, {Yocum}, {York}, {Zehavi}, {Zibetti}, \&
  {Zucker}}]{2005AJ....129.1755A}
---. 2005, \aj, 129, 1755

\bibitem[{{Adelman-McCarthy} {et~al.}(2006){Adelman-McCarthy}, {Ag{\"u}eros},
  {Allam}, {Anderson}, {Anderson}, {Annis}, {Bahcall}, {Baldry}, {Barentine},
  {Berlind}, {Bernardi}, {Blanton}, {Boroski}, {Brewington}, {Brinchmann},
  {Brinkmann}, {Brunner}, {Budav{\'a}ri}, {Carey}, {Carr}, {Castander},
  {Connolly}, {Csabai}, {Czarapata}, {Dalcanton}, {Doi}, {Dong}, {Eisenstein},
  {Evans}, {Fan}, {Finkbeiner}, {Friedman}, {Frieman}, {Fukugita}, {Gillespie},
  {Glazebrook}, {Gray}, {Grebel}, {Gunn}, {Gurbani}, {de Haas}, {Hall},
  {Harris}, {Harvanek}, {Hawley}, {Hayes}, {Hendry}, {Hennessy}, {Hindsley},
  {Hirata}, {Hogan}, {Hogg}, {Holmgren}, {Holtzman}, {Ichikawa}, {Ivezi{\'c}},
  {Jester}, {Johnston}, {Jorgensen}, {Juri{\'c}}, {Kent}, {Kleinman}, {Knapp},
  {Kniazev}, {Kron}, {Krzesinski}, {Kuropatkin}, {Lamb}, {Lampeitl}, {Lee},
  {Leger}, {Lin}, {Long}, {Loveday}, {Lupton}, {Margon},
  {Mart{\'{\i}}nez-Delgado}, {Mandelbaum}, {Matsubara}, {McGehee}, {McKay},
  {Meiksin}, {Munn}, {Nakajima}, {Nash}, {Neilsen}, {Newberg}, {Newman},
  {Nichol}, {Nicinski}, {Nieto-Santisteban}, {Nitta}, {O'Mullane}, {Okamura},
  {Owen}, {Padmanabhan}, {Pauls}, {Peoples}, {Pier}, {Pope}, {Pourbaix},
  {Quinn}, {Richards}, {Richmond}, {Rockosi}, {Schlegel}, {Schneider},
  {Schroeder}, {Scranton}, {Seljak}, {Sheldon}, {Shimasaku}, {Smith}, {Smol{\v
  c}i{\'c}}, {Snedden}, {Stoughton}, {Strauss}, {SubbaRao}, {Szalay},
  {Szapudi}, {Szkody}, {Tegmark}, {Thakar}, {Tucker}, {Uomoto}, {Vanden Berk},
  {Vandenberg}, {Vogeley}, {Voges}, {Vogt}, {Walkowicz}, {Weinberg}, {West},
  {White}, {Xu}, {Yanny}, {Yocum}, {York}, {Zehavi}, {Zibetti}, \&
  {Zucker}}]{2006ApJS..162...38A}
{Adelman-McCarthy}, J.~K. {et~al.} 2006, \apjs, 162, 38

\bibitem[{{Annis} {et~al.}(1999){Annis}, {Kent}, {Castander}, {Eisenstein},
  {Gunn}, {Kim}, {Lupton}, {Nichol}, {Postman}, {Voges}, \& {SDSS
  Collaboration}}]{1999AAS...195.1202A}
{Annis}, J. {et~al.} 1999, in Bulletin of the American Astronomical Society,
  1391--+

\bibitem[{{Bahcall} {et~al.}(2003){Bahcall}, {McKay}, {Annis}, {Kim}, {Dong},
  {Hansen}, {Goto}, {Gunn}, {Miller}, {Nichol}, {Postman}, {Schneider},
  {Schroeder}, {Voges}, {Brinkmann}, \& {Fukugita}}]{2003ApJS..148..243B}
{Bahcall}, N.~A. {et~al.} 2003, \apjs, 148, 243

\bibitem[{{Beers} {et~al.}(1990){Beers}, {Flynn}, \&
  {Gebhardt}}]{1990AJ....100...32B}
{Beers}, T.~C., {Flynn}, K., \& {Gebhardt}, K. 1990, \aj, 100, 32

\bibitem[{{Berlind} {et~al.}(2006){Berlind}, {Frieman}, {Weinberg}, {Blanton},
  {Warren}, {Abazajian}, {Scranton}, {Hogg}, {Scoccimarro}, {Bahcall},
  {Brinkmann}, {Gott}, {Kleinman}, {Krzesinski}, {Lee}, {Miller}, {Nitta},
  {Schneider}, {Tucker}, \& {Zehavi}}]{2006ApJS..167....1B}
{Berlind}, A.~A. {et~al.} 2006, \apjs, 167, 1

\bibitem[{{Biviano} {et~al.}(1992){Biviano}, {Girardi}, {Giuricin},
  {Mardirossian}, \& {Mezzetti}}]{1992ApJ...396...35B}
{Biviano}, A., {Girardi}, M., {Giuricin}, G., {Mardirossian}, F., \&
  {Mezzetti}, M. 1992, \apj, 396, 35

\bibitem[{{Biviano} \& {Girardi}(2003)}]{2003ApJ...585..205B}
{Biviano}, A., \& {Girardi}, M. 2003, \apj, 585, 205

\bibitem[{{Blanton} {et~al.}(2003){Blanton}, {Hogg}, {Bahcall}, {Brinkmann},
  {Britton}, {Connolly}, {Csabai}, {Fukugita}, {Loveday}, {Meiksin}, {Munn},
  {Nichol}, {Okamura}, {Quinn}, {Schneider}, {Shimasaku}, {Strauss}, {Tegmark},
  {Vogeley}, \& {Weinberg}}]{2003ApJ...592..819B}
{Blanton}, M.~R. {et~al.} 2003, \apj, 592, 819

\bibitem[{{Blanton} {et~al.}(2005){Blanton}, {Schlegel}, {Strauss},
  {Brinkmann}, {Finkbeiner}, {Fukugita}, {Gunn}, {Hogg}, {Ivezi{\'c}}, {Knapp},
  {Lupton}, {Munn}, {Schneider}, {Tegmark}, \& {Zehavi}}]{2005AJ....129.2562B}
---. 2005, \aj, 129, 2562

\bibitem[{{B{\"o}hringer} {et~al.}(2000){B{\"o}hringer}, {Voges}, {Huchra},
  {McLean}, {Giacconi}, {Rosati}, {Burg}, {Mader}, {Schuecker}, {Simi{\c c}},
  {Komossa}, {Reiprich}, {Retzlaff}, \& {Tr{\"u}mper}}]{2000ApJS..129..435B}
{B{\"o}hringer}, H. {et~al.} 2000, \apjs, 129, 435

\bibitem[{{Brainerd} \& {Specian}(2003)}]{2003ApJ...593L...7B}
{Brainerd}, T.~G., \& {Specian}, M.~A. 2003, \apjl, 593, L7

\bibitem[{{Buote} {et~al.}(2006){Buote}, {Gastaldello}, {Humphrey},
  {Zappacosta}, {Bullock}, {Brighenti}, \& {Mathews}}]{2006astro.ph.10135B}
{Buote}, D.~A., {Gastaldello}, F., {Humphrey}, P.~J., {Zappacosta}, L.,
  {Bullock}, J.~S., {Brighenti}, F., \& {Mathews}, W.~G. 2006, \apj, in press,
  arXiv:astro-ph/0610135

\bibitem[{{Caldwell}(1987)}]{1987AJ.....94.1116C}
{Caldwell}, N. 1987, \aj, 94, 1116

\bibitem[{{Carlberg}(1994)}]{1994ApJ...434L..51C}
{Carlberg}, R.~G. 1994, \apjl, 434, L51

\bibitem[{{Carlberg} {et~al.}(1997){Carlberg}, {Yee}, \&
  {Ellingson}}]{1997ApJ...478..462C}
{Carlberg}, R.~G., {Yee}, H.~K.~C., \& {Ellingson}, E. 1997, \apj, 478, 462

\bibitem[{{Col{\'{\i}}n} {et~al.}(2000){Col{\'{\i}}n}, {Klypin}, \&
  {Kravtsov}}]{2000ApJ...539..561C}
{Col{\'{\i}}n}, P., {Klypin}, A.~A., \& {Kravtsov}, A.~V. 2000, \apj, 539, 561

\bibitem[{{Connolly} {et~al.}(2000){Connolly}, {Genovese}, {Moore}, {Nichol},
  {Schneider}, \& {Wasserman}}]{2000astro.ph..8187C}
{Connolly}, A.~J., {Genovese}, C., {Moore}, A.~W., {Nichol}, R.~C.,
  {Schneider}, J., \& {Wasserman}, L. 2000, arXiv:astro-ph/0008187

\bibitem[{{Conroy} {et~al.}(2005){Conroy}, {Newman}, {Davis}, {Coil}, {Yan},
  {Cooper}, {Gerke}, {Faber}, \& {Koo}}]{2005ApJ...635..982C}
{Conroy}, C. {et~al.} 2005, \apj, 635, 982

\bibitem[{{Conroy} {et~al.}(2007){Conroy}, {Prada}, {Newman}, {Croton}, {Coil},
  {Conselice}, {Cooper}, {Davis}, {Faber}, {Gerke}, {Guhathakurta}, {Klypin},
  {Koo}, \& {Yan}}]{2007ApJ...654..153C}
---. 2007, \apj, 654, 153

\bibitem[{{Conroy} {et~al.}(2006){Conroy}, {Wechsler}, \&
  {Kravtsov}}]{2006ApJ...647..201C}
{Conroy}, C., {Wechsler}, R.~H., \& {Kravtsov}, A.~V. 2006, \apj, 647, 201

\bibitem[{{Cortese} {et~al.}(2004){Cortese}, {Gavazzi}, {Boselli},
  {Iglesias-Paramo}, \& {Carrasco}}]{2004A&A...425..429C}
{Cortese}, L., {Gavazzi}, G., {Boselli}, A., {Iglesias-Paramo}, J., \&
  {Carrasco}, L. 2004, \aap, 425, 429

\bibitem[{{Croton} {et~al.}(2007){Croton}, {Gao}, \&
  {White}}]{2007MNRAS.374.1303C}
{Croton}, D.~J., {Gao}, L., \& {White}, S.~D.~M. 2007, \mnras, 374, 1303

\bibitem[{{Dahle}(2006)}]{2006ApJ...653..954D}
{Dahle}, H. 2006, \apj, 653, 954

\bibitem[{Dempster {et~al.}(1977)Dempster, Laird, \& Rubin}]{dm77}
Dempster, A., Laird, N., \& Rubin, D. 1977, J. Roy. Stat. Soc., 39, 1

\bibitem[{{Diaferio}(1999)}]{1999MNRAS.309..610D}
{Diaferio}, A. 1999, \mnras, 309, 610

\bibitem[{{Diaferio} \& {Geller}(1996)}]{1996ApJ...467...19D}
{Diaferio}, A., \& {Geller}, M.~J. 1996, \apj, 467, 19

\bibitem[{{Diaferio} \& {Geller}(1997)}]{1997ApJ...481..633D}
---. 1997, \apj, 481, 633

\bibitem[{{Diaferio} {et~al.}(2005){Diaferio}, {Geller}, \&
  {Rines}}]{2005ApJ...628L..97D}
{Diaferio}, A., {Geller}, M.~J., \& {Rines}, K.~J. 2005, \apjl, 628, L97

\bibitem[{{Diemand} {et~al.}(2004){Diemand}, {Moore}, \&
  {Stadel}}]{2004MNRAS.352..535D}
{Diemand}, J., {Moore}, B., \& {Stadel}, J. 2004, \mnras, 352, 535

\bibitem[{{Eisenstein} {et~al.}(2001){Eisenstein}, {Annis}, {Gunn}, {Szalay},
  {Connolly}, {Nichol}, {Bahcall}, {Bernardi}, {Burles}, {Castander},
  {Fukugita}, {Hogg}, {Ivezi{\'c}}, {Knapp}, {Lupton}, {Narayanan}, {Postman},
  {Reichart}, {Richmond}, {Schneider}, {Schlegel}, {Strauss}, {SubbaRao},
  {Tucker}, {Vanden Berk}, {Vogeley}, {Weinberg}, \&
  {Yanny}}]{2001AJ....122.2267E}
{Eisenstein}, D.~J. {et~al.} 2001, \aj, 122, 2267

\bibitem[{{Evrard} {et~al.}(2007){Evrard}, {Bialek}, {Busha}, {White}, {Habib},
  {Heitmann}, {Warren}, {Rasia}, {Tormen}, {Moscardini}, {Power}, {Jenkins},
  {Gao}, {Frenk}, {Springel}, {White}, \& {Diemand}}]{2007astro.ph..2241E}
{Evrard}, A.~E. {et~al.} 2007, \apj, submitted

\bibitem[{{Evrard} {et~al.}(2002){Evrard}, {MacFarland}, {Couchman}, {Colberg},
  {Yoshida}, {White}, {Jenkins}, {Frenk}, {Pearce}, {Peacock}, \&
  {Thomas}}]{2002ApJ...573....7E}
---. 2002, \apj, 573, 7

\bibitem[{{Faltenbacher} \& {Diemand}(2006)}]{2006MNRAS.369.1698F}
{Faltenbacher}, A., \& {Diemand}, J. 2006, \mnras, 369, 1698

\bibitem[{{Faltenbacher} {et~al.}(2005){Faltenbacher}, {Kravtsov}, {Nagai}, \&
  {Gottl{\"o}ber}}]{2005MNRAS.358..139F}
{Faltenbacher}, A., {Kravtsov}, A.~V., {Nagai}, D., \& {Gottl{\"o}ber}, S.
  2005, \mnras, 358, 139

\bibitem[{{Frenk} {et~al.}(1996){Frenk}, {Evrard}, {White}, \&
  {Summers}}]{1996ApJ...472..460F}
{Frenk}, C.~S., {Evrard}, A.~E., {White}, S.~D.~M., \& {Summers}, F.~J. 1996,
  \apj, 472, 460

\bibitem[{{Fukugita} {et~al.}(1996){Fukugita}, {Ichikawa}, {Gunn}, {Doi},
  {Shimasaku}, \& {Schneider}}]{1996AJ....111.1748F}
{Fukugita}, M., {Ichikawa}, T., {Gunn}, J.~E., {Doi}, M., {Shimasaku}, K., \&
  {Schneider}, D.~P. 1996, \aj, 111, 1748

\bibitem[{{Gerke} {et~al.}(2005){Gerke}, {Newman}, {Davis}, {Marinoni}, {Yan},
  {Coil}, {Conroy}, {Cooper}, {Faber}, {Finkbeiner}, {Guhathakurta}, {Kaiser},
  {Koo}, {Phillips}, {Weiner}, \& {Willmer}}]{2005ApJ...625....6G}
{Gerke}, B.~F. {et~al.} 2005, \apj, 625, 6

\bibitem[{{Ghigna} {et~al.}(2000){Ghigna}, {Moore}, {Governato}, {Lake},
  {Quinn}, \& {Stadel}}]{2000ApJ...544..616G}
{Ghigna}, S., {Moore}, B., {Governato}, F., {Lake}, G., {Quinn}, T., \&
  {Stadel}, J. 2000, \apj, 544, 616

\bibitem[{{Girardi} {et~al.}(1993){Girardi}, {Biviano}, {Giuricin},
  {Mardirossian}, \& {Mezzetti}}]{1993ApJ...404...38G}
{Girardi}, M., {Biviano}, A., {Giuricin}, G., {Mardirossian}, F., \&
  {Mezzetti}, M. 1993, \apj, 404, 38

\bibitem[{{Girardi} {et~al.}(2005){Girardi}, {Demarco}, {Rosati}, \&
  {Borgani}}]{2005A&A...442...29G}
{Girardi}, M., {Demarco}, R., {Rosati}, P., \& {Borgani}, S. 2005, \aap, 442,
  29

\bibitem[{{Girardi} {et~al.}(1998){Girardi}, {Giuricin}, {Mardirossian},
  {Mezzetti}, \& {Boschin}}]{1998ApJ...505...74G}
{Girardi}, M., {Giuricin}, G., {Mardirossian}, F., {Mezzetti}, M., \&
  {Boschin}, W. 1998, \apj, 505, 74

\bibitem[{{Gladders} \& {Yee}(2000)}]{2000AJ....120.2148G}
{Gladders}, M.~D., \& {Yee}, H.~K.~C. 2000, \aj, 120, 2148

\bibitem[{{Gladders} \& {Yee}(2005)}]{2005ApJS..157....1G}
---. 2005, \apjs, 157, 1

\bibitem[{{Gladders} {et~al.}(2007){Gladders}, {Yee}, {Majumdar}, {Barrientos},
  {Hoekstra}, {Hall}, \& {Infante}}]{2007ApJ...655..128G}
{Gladders}, M.~D., {Yee}, H.~K.~C., {Majumdar}, S., {Barrientos}, L.~F.,
  {Hoekstra}, H., {Hall}, P.~B., \& {Infante}, L. 2007, \apj, 655, 128

\bibitem[{{Grego} {et~al.}(2000){Grego}, {Carlstrom}, {Joy}, {Reese}, {Holder},
  {Patel}, {Cooray}, \& {Holzapfel}}]{2000ApJ...539...39G}
{Grego}, L., {Carlstrom}, J.~E., {Joy}, M.~K., {Reese}, E.~D., {Holder}, G.~P.,
  {Patel}, S., {Cooray}, A.~R., \& {Holzapfel}, W.~L. 2000, \apj, 539, 39

\bibitem[{{Gunn} {et~al.}(1998){Gunn}, {Carr}, {Rockosi}, {Sekiguchi}, {Berry},
  {Elms}, {de Haas}, {Ivezi{\'c}}, {Knapp}, {Lupton}, {Pauls}, {Simcoe},
  {Hirsch}, {Sanford}, {Wang}, {York}, {Harris}, {Annis}, {Bartozek},
  {Boroski}, {Bakken}, {Haldeman}, {Kent}, {Holm}, {Holmgren}, {Petravick},
  {Prosapio}, {Rechenmacher}, {Doi}, {Fukugita}, {Shimasaku}, {Okada}, {Hull},
  {Siegmund}, {Mannery}, {Blouke}, {Heidtman}, {Schneider}, {Lucinio}, \&
  {Brinkman}}]{1998AJ....116.3040G}
{Gunn}, J.~E. {et~al.} 1998, \aj, 116, 3040

\bibitem[{{Halliday} {et~al.}(2004){Halliday}, {Milvang-Jensen}, {Poirier},
  {Poggianti}, {Jablonka}, {Arag{\'o}n-Salamanca}, {Saglia}, {De Lucia},
  {Pell{\'o}}, {Simard}, {Clowe}, {Rudnick}, {Dalcanton}, {White}, \&
  {Zaritsky}}]{2004A&A...427..397H}
{Halliday}, C. {et~al.} 2004, \aap, 427, 397

\bibitem[{{Hansen} {et~al.}(2005){Hansen}, {McKay}, {Wechsler}, {Annis},
  {Sheldon}, \& {Kimball}}]{2005ApJ...633..122H}
{Hansen}, S.~M., {McKay}, T.~A., {Wechsler}, R.~H., {Annis}, J., {Sheldon},
  E.~S., \& {Kimball}, A. 2005, \apj, 633, 122

\bibitem[{{Heisler} {et~al.}(1985){Heisler}, {Tremaine}, \&
  {Bahcall}}]{1985ApJ...298....8H}
{Heisler}, J., {Tremaine}, S., \& {Bahcall}, J.~N. 1985, \apj, 298, 8

\bibitem[{{Iguchi} {et~al.}(2005){Iguchi}, {Sota}, {Tatekawa}, {Nakamichi}, \&
  {Morikawa}}]{2005PhRvE..71a6102I}
{Iguchi}, O., {Sota}, Y., {Tatekawa}, T., {Nakamichi}, A., \& {Morikawa}, M.
  2005, \pre, 71, 016102

\bibitem[{{Izenman}(1991)}]{Izenman1991}
{Izenman}, A.~J. 1991, Journal of the American Statistical
Association, 86, 205

\bibitem[{{Jenkins} {et~al.}(2001){Jenkins}, {Frenk}, {White}, {Colberg},
  {Cole}, {Evrard}, {Couchman}, \& {Yoshida}}]{2001MNRAS.321..372J}
{Jenkins}, A., {Frenk}, C.~S., {White}, S.~D.~M., {Colberg}, J.~M., {Cole}, S.,
  {Evrard}, A.~E., {Couchman}, H.~M.~P., \& {Yoshida}, N. 2001, \mnras, 321,
  372

\bibitem[{{Johnston} {et~al.}(2007){Johnston}, {Sheldon},
    {Tasitsiomi}, {Frieman}, {Wechsler}, \&
    {McKay}}]{2007ApJ...656...27J}
{Johnston}, D.~E., {Sheldon}, E.~S., {Tasitsiomi}, A., {Frieman},
    J.~A., {Wechsler}, R.~H., \& {McKay}, T.~A. 2007, \apj, 656, 27

\bibitem[{{Katgert} {et~al.}(2004){Katgert}, {Biviano}, \&
  {Mazure}}]{2004ApJ...600..657K}
{Katgert}, P., {Biviano}, A., \& {Mazure}, A. 2004, \apj, 600, 657

\bibitem[{{Koester} {et~al.}(2007{\natexlab{a}}){Koester}, {McKay}, {Annis},
  {Wechsler}, {Evrard}, {Bleem}, {Becker}, {Johnston}, {Sheldon}, {Nichol},
  {Miller}, {Scranton}, {Bahcall}, {Barentine}, {Brewington}, {Brinkmann},
  {Harvanek}, {Kleinman}, {Krzesinski}, {Long}, {Nitta}, {Schneider},
  {Sneddin}, {Voges}, {York}, \& {SDSS collaboration}}]{2007astro.ph..1265K}
{Koester}, B.~P. {et~al.} 2007{\natexlab{a}}, \apj, in press,
  arXiv:astro-ph/0701265

\bibitem[{{Koester} {et~al.}(2007{\natexlab{b}}){Koester}, {McKay}, {Annis},
  {Wechsler}, {Evrard}, {Rozo}, {Bleem}, {Sheldon}, \&
  {Johnston}}]{2007astro.ph..1268K}
---. 2007{\natexlab{b}}, \apj, in press, arXiv:astro-ph/0701268

\bibitem[{{Kravtsov} {et~al.}(2004){Kravtsov}, {Berlind}, {Wechsler}, {Klypin},
  {Gottl{\"o}ber}, {Allgood}, \& {Primack}}]{2004ApJ...609...35K}
{Kravtsov}, A.~V., {Berlind}, A.~A., {Wechsler}, R.~H., {Klypin}, A.~A.,
  {Gottl{\"o}ber}, S., {Allgood}, B., \& {Primack}, J.~R. 2004, \apj, 609, 35

\bibitem[{{Lancaster} {et~al.}(2005){Lancaster}, {Genova-Santos}, {Falc{\`o}n},
  {Grainge}, {Guti{\`e}rrez}, {Kneissl}, {Marshall}, {Pooley}, {Rebolo},
  {Rubi{\~n}o-Martin}, {Saunders}, {Waldram}, \&
  {Watson}}]{2005MNRAS.359...16L}
{Lancaster}, K. {et~al.} 2005, \mnras, 359, 16

\bibitem[{{Levine} {et~al.}(2002){Levine}, {Schulz}, \&
  {White}}]{2002ApJ...577..569L}
{Levine}, E.~S., {Schulz}, A.~E., \& {White}, M. 2002, \apj, 577, 569

\bibitem[{{Lima} \& {Hu}(2004)}]{2004PhRvD..70d3504L}
{Lima}, M., \& {Hu}, W. 2004, \prd, 70, 043504

\bibitem[{{Lima} \& {Hu}(2005)}]{2005PhRvD..72d3006L}
---. 2005, \prd, 72, 043006

\bibitem[{{Mazure} {et~al.}(1996){Mazure}, {Katgert}, {den Hartog}, {Biviano},
  {Dubath}, {Escalera}, {Focardi}, {Gerbal}, {Giuricin}, {Jones}, {Le Fevre},
  {Moles}, {Perea}, \& {Rhee}}]{1996A&A...310...31M}
{Mazure}, A. {et~al.} 1996, \aap, 310, 31

\bibitem[{{McKay} {et~al.}(2002){McKay}, {Sheldon}, {Johnston}, {Grebel},
  {Prada}, {Rix}, {Bahcall}, {Brinkmann}, {Csabai}, {Fukugita}, {Lamb}, \&
  {York}}]{2002ApJ...571L..85M}
{McKay}, T.~A. {et~al.} 2002, \apjl, 571, L85

\bibitem[{{Miller} {et~al.}(2005){Miller}, {Nichol}, {Reichart}, {Wechsler},
  {Evrard}, {Annis}, {McKay}, {Bahcall}, {Bernardi}, {Boehringer}, {Connolly},
  {Goto}, {Kniazev}, {Lamb}, {Postman}, {Schneider}, {Sheth}, \&
  {Voges}}]{2005AJ....130..968M}
{Miller}, C.~J. {et~al.} 2005, \aj, 130, 968

\bibitem[{Navarro} {et~al.}(1997){Navarro}, {Frenk}, \&
  {White}]{1997ApJ...490..493N}
{Navarro}, J.~F., {Frenk}, C.~S., \& {White}, S.~D.~M. 1997, \apj,
  490, 493

\bibitem[{{Popesso} {et~al.}(2004){Popesso}, {Boehringer}, \&
  {Voges}}]{2004astro.ph..3357P}
{Popesso}, P., {Boehringer}, H., \& {Voges}, W. 2004, arXiv:astro-ph/0403357

\bibitem[{{Prada} {et~al.}(2003){Prada}, {Vitvitska}, {Klypin}, {Holtzman},
  {Schlegel}, {Grebel}, {Rix}, {Brinkmann}, {McKay}, \&
  {Csabai}}]{2003ApJ...598..260P}
{Prada}, F. {et~al.} 2003, \apj, 598, 260

\bibitem[{{Rines} \& {Diaferio}(2006)}]{2006AJ....132.1275R}
{Rines}, K., \& {Diaferio}, A. 2006, \aj, 132, 1275

\bibitem[{{Rines} {et~al.}(2006){Rines}, {Diaferio}, \&
  {Natarajan}}]{2006astro.ph..6545R}
{Rines}, K., {Diaferio}, A., \& {Natarajan}, P. 2006, \apj, in press,
  arXiv:astro-ph/0606545

\bibitem[{{Rines} {et~al.}(2003){Rines}, {Geller}, {Kurtz}, \&
  {Diaferio}}]{2003AJ....126.2152R}
{Rines}, K., {Geller}, M.~J., {Kurtz}, M.~J., \& {Diaferio}, A. 2003, \aj, 126,
  2152

\bibitem[{{Rosati} {et~al.}(1998){Rosati}, {della Ceca}, {Norman}, \&
  {Giacconi}}]{1998ApJ...492L..21R}
{Rosati}, P., {della Ceca}, R., {Norman}, C., \& {Giacconi}, R. 1998, \apjl,
  492, L21+

\bibitem[{{Rozo} {et~al.}(2007{\natexlab{a}}){Rozo}, {Wechsler}, {Koester},
  {Evrard}, \& {McKay}}]{2007astro.ph..3574R}
{Rozo}, E., {Wechsler}, R.~H., {Koester}, B.~P., {Evrard}, A.~E., \& {McKay},
  T.~A. 2007{\natexlab{a}}, \apj, submitted, astro-ph/0703574

\bibitem[{{Rozo} {et~al.}(2007{\natexlab{b}}){Rozo}, {Wechsler}, {Koester},
  {McKay}, {Evrard}, {Johnston}, {Sheldon}, {Annis}, \&
  {Frieman}}]{2007astro.ph..3571R}
{Rozo}, E. {et~al.} 2007{\natexlab{b}}, \apj, submitted, astro-ph/0703571

\bibitem[{{Sheth}(1996)}]{1996MNRAS.279.1310S}
{Sheth}, R.~K. 1996, \mnras, 279, 1310

\bibitem[{{Sheth} {et~al.}(2003){Sheth}, {Bernardi}, {Schechter}, {Burles},
  {Eisenstein}, {Finkbeiner}, {Frieman}, {Lupton}, {Schlegel}, {Subbarao},
  {Shimasaku}, {Bahcall}, {Brinkmann}, \& {Ivezi{\'c}}}]{2003ApJ...594..225S}
{Sheth}, R.~K. {et~al.} 2003, \apj, 594, 225

\bibitem[{{Sheth} \& {Diaferio}(2001)}]{2001MNRAS.322..901S}
{Sheth}, R.~K., \& {Diaferio}, A. 2001, \mnras, 322, 901

\bibitem[{{Smith} {et~al.}(2002){Smith}, {Tucker}, {Kent}, {Richmond},
  {Fukugita}, {Ichikawa}, {Ichikawa}, {Jorgensen}, {Uomoto}, {Gunn}, {Hamabe},
  {Watanabe}, {Tolea}, {Henden}, {Annis}, {Pier}, {McKay}, {Brinkmann}, {Chen},
  {Holtzman}, {Shimasaku}, \& {York}}]{2002AJ....123.2121S}
{Smith}, J.~A. {et~al.} 2002, \aj, 123, 2121

\bibitem[{{Smith}(1936)}]{1936ApJ....83...23S}
{Smith}, S. 1936, \apj, 83, 23

\bibitem[{{Spergel} {et~al.}(2006){Spergel}, {Bean}, {Dor{\'e}}, {Nolta},
  {Bennett}, {Dunkley}, {Hinshaw}, {Jarosik}, {Komatsu}, {Page}, {Peiris},
  {Verde}, {Halpern}, {Hill}, {Kogut}, {Limon}, {Meyer}, {Odegard}, {Tucker},
  {Weiland}, {Wollack}, \& {Wright}}]{2006astro.ph..3449S}
{Spergel}, D.~N. {et~al.} 2006, \apj, in press, arXiv:astro-ph/0603449

\bibitem[{{Springel} {et~al.}(2005){Springel}, {White}, {Jenkins}, {Frenk},
  {Yoshida}, {Gao}, {Navarro}, {Thacker}, {Croton}, {Helly}, {Peacock}, {Cole},
  {Thomas}, {Couchman}, {Evrard}, {Colberg}, \& {Pearce}}]{2005Natur.435..629S}
{Springel}, V. {et~al.} 2005, \nat, 435, 629

\bibitem[{{Stanek} {et~al.}(2006){Stanek}, {Evrard}, {B{\"o}hringer},
  {Schuecker}, \& {Nord}}]{2006ApJ...648..956S}
{Stanek}, R., {Evrard}, A.~E., {B{\"o}hringer}, H., {Schuecker}, P., \& {Nord},
  B. 2006, \apj, 648, 956

\bibitem[{{Strauss} {et~al.}(2002){Strauss}, {Weinberg}, {Lupton}, {Narayanan},
  {Annis}, {Bernardi}, {Blanton}, {Burles}, {Connolly}, {Dalcanton}, {Doi},
  {Eisenstein}, {Frieman}, {Fukugita}, {Gunn}, {Ivezi{\'c}}, {Kent}, {Kim},
  {Knapp}, {Kron}, {Munn}, {Newberg}, {Nichol}, {Okamura}, {Quinn}, {Richmond},
  {Schlegel}, {Shimasaku}, {SubbaRao}, {Szalay}, {Vanden Berk}, {Vogeley},
  {Yanny}, {Yasuda}, {York}, \& {Zehavi}}]{2002AJ....124.1810S}
{Strauss}, M.~A. {et~al.} 2002, \aj, 124, 1810

\bibitem[{{Struble} \& {Rood}(1999)}]{1999ApJS..125...35S}
{Struble}, M.~F., \& {Rood}, H.~J. 1999, \apjs, 125, 35

\bibitem[{{Tasitsiomi} {et~al.}(2004){Tasitsiomi}, {Kravtsov}, {Wechsler}, \&
  {Primack}}]{2004ApJ...614..533T}
{Tasitsiomi}, A., {Kravtsov}, A.~V., {Wechsler}, R.~H., \& {Primack}, J.~R.
  2004, \apj, 614, 533

\bibitem[{{Tegmark} {et~al.}(2006){Tegmark}, {Eisenstein}, {Strauss},
  {Weinberg}, {Blanton}, {Frieman}, {Fukugita}, {Gunn}, {Hamilton}, {Knapp},
  {Nichol}, {Ostriker}, {Padmanabhan}, {Percival}, {Schlegel}, {Schneider},
  {Scoccimarro}, {Seljak}, {Seo}, {Swanson}, {Szalay}, {Vogeley}, {Yoo},
  {Zehavi}, {Abazajian}, {Anderson}, {Annis}, {Bahcall}, {Bassett}, {Berlind},
  {Brinkmann}, {Budavari}, {Castander}, {Connolly}, {Csabai}, {Doi},
  {Finkbeiner}, {Gillespie}, {Glazebrook}, {Hennessy}, {Hogg}, {Ivezi{\'c}},
  {Jain}, {Johnston}, {Kent}, {Lamb}, {Lee}, {Lin}, {Loveday}, {Lupton},
  {Munn}, {Pan}, {Park}, {Peoples}, {Pier}, {Pope}, {Richmond}, {Rockosi},
  {Scranton}, {Sheth}, {Stebbins}, {Stoughton}, {Szapudi}, {Tucker}, {Berk},
  {Yanny}, \& {York}}]{2006PhRvD..74l3507T}
{Tegmark}, M. {et~al.} 2006, \prd, 74, 123507

\bibitem[{{van den Bosch} {et~al.}(2004){van den Bosch}, {Norberg}, {Mo}, \&
  {Yang}}]{2004MNRAS.352.1302V}
{van den Bosch}, F.~C., {Norberg}, P., {Mo}, H.~J., \& {Yang}, X. 2004, \mnras,
  352, 1302

\bibitem[{{van den Bosch} {et~al.}(2005){van den Bosch}, {Weinmann}, {Yang},
  {Mo}, {Li}, \& {Jing}}]{2005MNRAS.361.1203V}
{van den Bosch}, F.~C., {Weinmann}, S.~M., {Yang}, X., {Mo}, H.~J., {Li}, C.,
  \& {Jing}, Y.~P. 2005, \mnras, 361, 1203

\bibitem[{{van der Marel} \& {Franx}(1993)}]{1993ApJ...407..525V}
{van der Marel}, R.~P., \& {Franx}, M. 1993, \apj, 407, 525

\bibitem[{{van der Marel} {et~al.}(2000){van der Marel}, {Magorrian},
  {Carlberg}, {Yee}, \& {Ellingson}}]{2000AJ....119.2038V}
{van der Marel}, R.~P., {Magorrian}, J., {Carlberg}, R.~G., {Yee}, H.~K.~C., \&
  {Ellingson}, E. 2000, \aj, 119, 2038

\bibitem[{{Warren}(1994)}]{1994PhDT.........1W}
{Warren}, M.~S. 1994, PhD thesis, AA(California Univ., Santa Barbara, CA.)

\bibitem[{{Wasserman} {et~al.}(2001){Wasserman}, {Miller}, {Nichol},
  {Genovese}, {Jang}, {Connolly}, {Moore}, {Schneider}, \& {the PICA
  group}}]{2001astro.ph.12050W}
{Wasserman}, L. {et~al.} 2001, arXiv:astro-ph/0112050

\bibitem[{{Wechsler} {et~al.}(2006){Wechsler}, {Zentner}, {Bullock},
  {Kravtsov}, \& {Allgood}}]{2006ApJ...652...71W}
{Wechsler}, R.~H., {Zentner}, A.~R., {Bullock}, J.~S., {Kravtsov}, A.~V., \&
  {Allgood}, B. 2006, \apj, 652, 71

\bibitem[{{Wittman} {et~al.}(2006){Wittman}, {Dell'Antonio}, {Hughes},
  {Margoniner}, {Tyson}, {Cohen}, \& {Norman}}]{2006ApJ...643..128W}
{Wittman}, D., {Dell'Antonio}, I.~P., {Hughes}, J.~P., {Margoniner}, V.~E.,
  {Tyson}, J.~A., {Cohen}, J.~G., \& {Norman}, D. 2006, \apj, 643, 128

\bibitem[{{Wojtak} {et~al.}(2006){Wojtak}, {Lokas}, {Mamon}, {Gottloeber},
  {Prada}, \& {Moles}}]{2006astro.ph..6579W}
{Wojtak}, R., {Lokas}, E.~L., {Mamon}, G.~A., {Gottloeber}, S., {Prada}, F., \&
  {Moles}, M. 2006, \aap, in press, arXiv:astro-ph/0606579

\bibitem[{{York} {et~al.}(2000){York}, {Adelman}, {Anderson}, {Anderson},
  {Annis}, {Bahcall}, {Bakken}, {Barkhouser}, {Bastian}, {Berman}, {Boroski},
  {Bracker}, {Briegel}, {Briggs}, {Brinkmann}, {Brunner}, {Burles}, {Carey},
  {Carr}, {Castander}, {Chen}, {Colestock}, {Connolly}, {Crocker}, {Csabai},
  {Czarapata}, {Davis}, {Doi}, {Dombeck}, {Eisenstein}, {Ellman}, {Elms},
  {Evans}, {Fan}, {Federwitz}, {Fiscelli}, {Friedman}, {Frieman}, {Fukugita},
  {Gillespie}, {Gunn}, {Gurbani}, {de Haas}, {Haldeman}, {Harris}, {Hayes},
  {Heckman}, {Hennessy}, {Hindsley}, {Holm}, {Holmgren}, {Huang}, {Hull},
  {Husby}, {Ichikawa}, {Ichikawa}, {Ivezi{\'c}}, {Kent}, {Kim}, {Kinney},
  {Klaene}, {Kleinman}, {Kleinman}, {Knapp}, {Korienek}, {Kron}, {Kunszt},
  {Lamb}, {Lee}, {Leger}, {Limmongkol}, {Lindenmeyer}, {Long}, {Loomis},
  {Loveday}, {Lucinio}, {Lupton}, {MacKinnon}, {Mannery}, {Mantsch}, {Margon},
  {McGehee}, {McKay}, {Meiksin}, {Merelli}, {Monet}, {Munn}, {Narayanan},
  {Nash}, {Neilsen}, {Neswold}, {Newberg}, {Nichol}, {Nicinski}, {Nonino},
  {Okada}, {Okamura}, {Ostriker}, {Owen}, {Pauls}, {Peoples}, {Peterson},
  {Petravick}, {Pier}, {Pope}, {Pordes}, {Prosapio}, {Rechenmacher}, {Quinn},
  {Richards}, {Richmond}, {Rivetta}, {Rockosi}, {Ruthmansdorfer}, {Sandford},
  {Schlegel}, {Schneider}, {Sekiguchi}, {Sergey}, {Shimasaku}, {Siegmund},
  {Smee}, {Smith}, {Snedden}, {Stone}, {Stoughton}, {Strauss}, {Stubbs},
  {SubbaRao}, {Szalay}, {Szapudi}, {Szokoly}, {Thakar}, {Tremonti}, {Tucker},
  {Uomoto}, {Vanden Berk}, {Vogeley}, {Waddell}, {Wang}, {Watanabe},
  {Weinberg}, {Yanny}, \& {Yasuda}}]{2000AJ....120.1579Y}
{York}, D.~G. {et~al.} 2000, \aj, 120, 1579

\bibitem[{{Zehavi} {et~al.}(2005){Zehavi}, {Zheng}, {Weinberg}, {Frieman},
  {Berlind}, {Blanton}, {Scoccimarro}, {Sheth}, {Strauss}, {Kayo}, {Suto},
  {Fukugita}, {Nakamura}, {Bahcall}, {Brinkmann}, {Gunn}, {Hennessy},
  {Ivezi{\'c}}, {Knapp}, {Loveday}, {Meiksin}, {Schlegel}, {Schneider},
  {Szapudi}, {Tegmark}, {Vogeley}, \& {York}}]{2005ApJ...630....1Z}
{Zehavi}, I. {et~al.} 2005, \apj, 630, 1

\bibitem[{{Zentner} {et~al.}(2005){Zentner}, {Berlind}, {Bullock}, {Kravtsov},
  \& {Wechsler}}]{2005ApJ...624..505Z}
{Zentner}, A.~R., {Berlind}, A.~A., {Bullock}, J.~S., {Kravtsov}, A.~V., \&
  {Wechsler}, R.~H. 2005, \apj, 624, 505

\bibitem[{{Zwicky}(1933)}]{1933AcHPh...6..110Z}
{Zwicky}, F. 1933, Helvetica Physica Acta, 6, 110

\bibitem[{{Zwicky}(1937)}]{1937ApJ....86..217Z}
---. 1937, \apj, 86, 217

\end{thebibliography}
\bibliographystyle{apj}




\end{document}